\documentclass[aps,prb,twocolumn,superscriptaddress,longbibliography,floatfix]{revtex4-2}
\usepackage{amsmath,amssymb,bm}
\usepackage{placeins}
\usepackage{graphicx}
\usepackage{microtype}
\usepackage{hyperref}
\hypersetup{colorlinks=true,linkcolor=blue,citecolor=blue,urlcolor=blue}

\begin{document}
\raggedbottom

\title{Inter-harmonic ratio structure and saturation of Bernstein modes in graphene}

\author{Miguel Tierz}
\email{tierz@simis.cn}
\affiliation{Shanghai Institute for Mathematics and Interdisciplinary Sciences (SIMIS), Shanghai 200438, China}
\date{May 6, 2026}

\begin{abstract}
Bernstein modes (BM) in graphene are finite-wavevector magnetoplasmons excited by contact near fields, whereas ordinary cyclotron resonance (CR) probes $q\approx0$. We derive the BM peak absorption in the quasiclassical ballistic regime and show that it factorizes into a launch spectrum, Bernstein-mode splitting, turning-point enhancement, and residual dielectric-response factor. At fixed excitation frequency, BM overtones ($n\ge2$) are sampled, to leading order, at the same momentum $q\simeq\omega/v_F$. Smooth launch and screening factors therefore cancel in inter-harmonic peak ratios, yielding $I_n/I_m\simeq m/n$, modified by linewidth corrections and one residual response ratio for each harmonic pair. In smooth-launcher synthetic tests, noisy full-$q$ spectra recover the residual ratio within errors: moderate launcher/dielectric misspecification within this benchmark family shifts it by only $\sim\!1$--$2\%$, whereas linewidth assumptions shift it by $\sim\!10$--$30\%$. The same factorization connects low-power amplitudes to nonlinear saturation. If BM harmonics share the same cooling region and bolometric readout, the low-power slope times onset intensity is harmonic independent, while BM and CR power sweeps obey distinct normalized saturation curves with linewidth scalings $\Gamma^{-1/2}$ and $\Gamma^{-1}$.
\end{abstract}

\maketitle

\section{Introduction}

Bernstein modes (BM) are collective charge excitations of a magnetized electron fluid. The two constituent effects are familiar separately: a magnetic field produces cyclotron motion, while a two-dimensional (2D) conductor supports plasmons, namely self-consistent oscillations of charge density and electric potential. BM physics appears when these effects meet at \emph{finite} in-plane wavevector $q$: the plasmon branch approaches an integer multiple $n\omega_c$ of the cyclotron frequency and hybridizes with it, producing the avoided-crossing structure first analyzed in kinetic plasma theory~\cite{SitenkoStepanov1957,Bernstein1958}. The quantum-mechanical magnetoplasmon dispersion of a two-dimensional electron gas (2DEG), including the Bernstein-mode branch structure, was subsequently obtained~\cite{ChiuQuinn1974}, and finite-$q$ geometric and grating-coupled BM/magnetoplasmon resonances were subsequently studied in semiconductor 2DEGs~\cite{batke1985nonlocality,ChaplikHeitmann1985,BatkeHeitmannTu1986,Lefebvre1998SST,Bangert1996,Holland2004PRL,Volkov2014PRB,Dorozhkin2021JETPL,Yavorskiy2025PRB}. In this sense, a Bernstein mode is not a separate elementary excitation but a plasmon--cyclotron hybrid that exists because the electronic response is nonlocal, i.e. depends on both frequency $\omega$ and wavevector $q$.

Graphene is a particularly suitable platform for this physics. It supports gate-tunable plasmons over a broad range of densities and frequencies~\cite{CastroNeto2009RMP,Grigorenko2012NatPhot}, and magnetoplasmon effects in the terahertz (THz) regime are well established~\cite{Crassee2012NanoLett,Yan2012NanoLett}. Plasmon-assisted resonant THz detection has also been demonstrated in graphene transistors~\cite{Bandurin2018NatComm}, and active electrical control of graphene plasmon damping has been achieved~\cite{ZhangFan2022PRL}. In doped graphene one may still use the quasiclassical cyclotron frequency $\omega_c=eB/m_c$, with $m_c=E_F/v_F^2$ the cyclotron mass~\cite{CastroNeto2009RMP,Goerbig2011RMP}, so the language of cyclotron harmonics remains natural even though the underlying band structure is Dirac-like. In the recent experiments, the relevant excitation is not the ordinary far-field $q\approx 0$ probe of cyclotron resonance (CR). Instead, the metallic contact launches an \emph{evanescent near field} containing a broad distribution of finite wavevectors $q$~\cite{Bandurin2022NatPhys}, and these finite-$q$ components couple efficiently to BM branches.

\begin{figure*}[!tbp]
\includegraphics[width=0.85\textwidth]{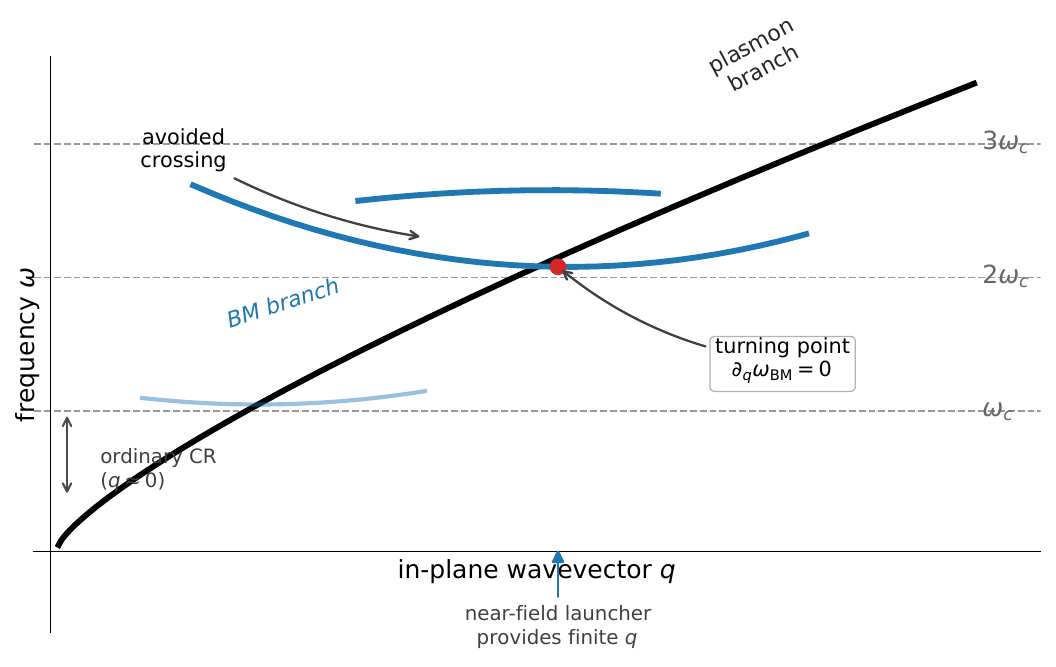}
\caption{Bernstein-mode geometry. The 2D plasmon branch (black) rises with in-plane wavevector $q$ and hybridizes with cyclotron harmonics $\omega=n\omega_c$ (gray dashed), producing avoided crossings. The lower Bernstein branch (blue) arches toward $n\omega_c$ from below and flattens at a turning point ($\partial_q\omega_{\rm BM}=0$, red dot), where the group velocity vanishes and absorption is strongly enhanced~\cite{Bandurin2022NatPhys}; the upper branch emerges above $n\omega_c$. Both branches possess turning points; the single-pole resonant form Eq.~\eqref{eq:epsL} describes the upper branch ($\omega=n\omega_c+\Omega_n>n\omega_c$), but the splitting $\Omega_n$, the curvature scaling, and the resulting $n^{-1}$ ordering of peak heights are common to both. Ordinary far-field cyclotron resonance probes $q\approx 0$; the contact near field supplies the finite-$q$ components needed to reach the turning-point region.}
\label{fig:bm-schematic}
\end{figure*}

Figure~\ref{fig:bm-schematic} summarizes the geometry. At fixed magnetic field, a rising plasmon branch meets the horizontal lines $\omega=n\omega_c$ and hybridizes with them at finite $q$, creating the avoided crossings at each integer harmonic. Close to a resonance, one of the hybridized branches can become locally very flat and develop a turning point, meaning $\partial_q\omega_{\rm BM}=0$~\cite{Roldan2011PRB}. That is precisely the regime emphasized in the near-field THz experiments~\cite{Bandurin2022NatPhys,Yahniuk2025}: a spatially uniform far-field probe would mainly access $q\approx 0$ and therefore ordinary CR, whereas the contact-generated evanescent field provides the finite momentum needed to reach the BM turning-point region.

A central geometric feature is the \emph{turning point} of the BM dispersion $\omega_{\rm BM}(q)$, meaning a point where the slope $\partial_q\omega_{\rm BM}$ vanishes~\cite{Roldan2011PRB,Kapralov2022PRB}. Near such a point the group velocity becomes small, the mode density is enhanced, and the absorption acquires the asymmetric square-root lineshape identified in experiment~\cite{Bandurin2022NatPhys}. The same turning-point structure is also responsible for the characteristic $\sqrt{\tau_p}$ scaling of the BM peak height~\cite{Bandurin2022NatPhys}, where $\tau_p$ is the momentum-relaxation time. In practical terms, the turning point is where a broad near-field momentum distribution is most efficiently converted into a large resonant signal. This geometric mechanism explains why BM peaks can exceed estimates based only on harmonic-overlap weights.

The theory of Bernstein modes in graphene was developed in Ref.~\cite{Roldan2011PRB}. A full kinetic model interpolating between ballistic and hydrodynamic magnetotransport was then constructed in Ref.~\cite{Kapralov2022PRB}: the avoided-crossing structure with its flat turning-point regions emerges only in the ballistic regime ($\omega\tau_{ee}\gg 1$), and formal long-wavelength expansions of the kinetic equations fail precisely at the turning-point wavevectors relevant here. The factorization developed below is therefore situated in the ballistic regime where the full nonlocal response, and in particular the Bessel-function harmonic structure, is essential. Experimentally, giant THz absorption peaks at the second and third cyclotron overtones were observed in Ref.~\cite{Bandurin2022NatPhys}, building on earlier THz-induced magneto-oscillation measurements~\cite{Monch2020NanoLett}. The BM square-root lineshape was subsequently used to extract angular-harmonic lifetimes~\cite{Moiseenko2025}, making it important to keep harmonic-dependent linewidths explicit in the inter-harmonic analysis derived below. Most recently, BM resonances were found to saturate at incident intensities well below those required for ordinary CR~\cite{Yahniuk2025}.

These developments leave two interpretive issues that remain incompletely separated. First, when one compares the amplitudes of different BM harmonics~\cite{Bandurin2022NatPhys,Yahniuk2025}, several effects are mixed together: the intrinsic harmonic structure of the electronic response~\cite{Dmitriev2012RMP,Kapralov2022PRB}, the momentum selectivity of the near-field launcher~\cite{Bandurin2022NatPhys}, gate screening~\cite{Hwang2007PRB}, the local curvature of the BM branch~\cite{Roldan2011PRB}, and harmonic-dependent linewidths, including angular electron-electron relaxation~\cite{Moiseenko2025}. Without separating those pieces, it is difficult to determine which amplitude ratios are universal and which are specific to a given device geometry. Second, the recent nonlinear data~\cite{Yahniuk2025} strongly suggest local hot-electron heating~\cite{BistrMac2009,SongReizer2012}, but the link between the \emph{linear} absorption coefficient and the observed saturation scale has not been written in a form that can be tested directly across harmonics.

More concretely, Refs.~\cite{Bandurin2022NatPhys,Yahniuk2025} established that BM peaks at several harmonics can be large, but not which part of their amplitude hierarchy is universal and which part is device-specific. Here we show that, at fixed excitation frequency, the common launcher dependence cancels at leading order from inter-harmonic BM ratios, leaving a parameter-free $m/n$ baseline together with a reduced response ratio, while harmonic-dependent linewidths of the type emphasized in Ref.~\cite{Moiseenko2025} enter as explicit corrections. We then connect this same low-power absorption coefficient to the nonlinear onset scales reported in Ref.~\cite{Yahniuk2025} through a minimal hot-electron energy-balance model.

The absolute electrodynamic version of this problem was already formulated in the Supplementary Information of Ref.~\cite{Bandurin2022NatPhys}: the contact produces a finite-$q$ diffraction spectrum, the graphene/substrate environment screens it, and the absorbed power is obtained by integrating the nonlocal loss function over the launched momenta. The present work does not attempt to replace that device-specific calculation. Instead, it asks which combinations of measured BM peak amplitudes are insensitive to the smooth part of the launcher and screening functions. The fixed-frequency inter-harmonic ratios derived below are precisely such combinations.

To fix notation, the integer $n$ labels the cyclotron \emph{harmonic} ($\omega\approx n\omega_c$), not a Landau level. The wavevector $q$ is supplied by the contact near field, and $R_c=v_F/\omega_c$ is the cyclotron radius~\cite{CastroNeto2009RMP}. The \emph{launcher spectrum} $D(q,\phi)$ measures how efficiently the contact excites wavevector $q$ for incident polarization angle $\phi$~\cite{Bandurin2022NatPhys}. Table~\ref{tab:notation} collects the remaining recurring symbols, grouped by their role in the argument.

\begin{table*}[!tbp]
\caption{Notation guide for the symbols used repeatedly in the main text. The table is intentionally selective: it lists the quantities that are most useful to keep in view while reading the factorized BM argument.}
\label{tab:notation}
\begin{ruledtabular}
\begin{tabular}{p{0.19\textwidth}p{0.75\textwidth}}
Symbol & Meaning and physical role \\
\hline
\multicolumn{2}{l}{\emph{General kinematics and response}} \\
$n$ & Cyclotron harmonic index, defined by $\omega\approx n\omega_c$; it is not a Landau-level index. \\
$q$ & In-plane wavevector launched by the contact; the near-field excitation is a distribution over $q$. \\
$R_c=v_F/\omega_c$ & Cyclotron radius. The turning-point condition is roughly $qR_c\simeq n$. \\
$q_n=n/R_c$ & Leading-order turning-point momentum for the $n$th harmonic; the physical turning point $q_n^*$ and Bessel-peak momentum $q_n^{(h)}$ differ from $q_n$ at the $\sim\!20$--$50\%$ level (Table~\ref{tab:qshift}). \\
$k_\omega=\omega/v_F$ & Common leading-order wavevector selected when different harmonics are compared at the same excitation frequency. \\
$\varepsilon(q,\omega)$ & Longitudinal dielectric function; the loss function $-\Im\,\varepsilon^{-1}(q,\omega)$ controls absorption. \\
\hline
\multicolumn{2}{l}{\emph{Factors in the factorized BM peak}} \\
$D(q,\phi)$ & Launcher spectrum: contact-generated field amplitude at wavevector $q$ for incident linear-polarization angle $\phi$~\cite{Bandurin2022NatPhys}. \\
$V(q)$ & Gate-screened Coulomb interaction entering the longitudinal dielectric response~\cite{Hwang2007PRB,Wunsch2006NJP}. \\
$J_n(qR_c)$ & Bessel-function harmonic form factor generated by semiclassical cyclotron motion~\cite{Dmitriev2012RMP,Bernstein1958}. \\
$\Gamma_n$ & Fitted BM linewidth in frequency units; in the simplest disorder-broadened model $\Gamma_n=\tau_p^{-1}$, but in the extraction protocol it is kept harmonic-dependent. \\
$\mathcal K_n$ & Local curvature $|\partial_q^2\Omega_n|$ at the turning point~\cite{Roldan2011PRB}; smaller $\mathcal K_n$ means a flatter branch and stronger turning-point enhancement. \\
$\mathcal G_n^{\rm int}$ & Intrinsic dielectric residue at the on-shell peak (Eq.~\eqref{eq:Gn-def}); the reduced peak factor of the loss function after explicit launcher, splitting, and turning-point factors are stripped. \\
$\mathcal G_n^{\rm eff}\equiv\beta_n\mathcal G_n^{\rm int}$ & Photoresistance-extracted effective residue; reduces to $\mathcal G_n^{\rm int}$ when the bolometric transfer $\beta_n$ is common across harmonics. \\
\hline
\multicolumn{2}{l}{\emph{Symbols used in the nonlinear model}} \\
$\alpha_{\nu,0},\ s_\nu$ & Low-power peak \emph{absorption} efficiency and linewidth exponent for resonance type $\nu$; $s_{\rm BM}=1/2$ (Eq.~\eqref{eq:exponents}) and $s_{\rm CR}=1$. \\
$A_n=\beta_n\alpha_{n,0}$ & Measured photoresistance slope; the bolometric transfer $\beta_n$ connects absorption to detection (Sec.~\ref{sec:abs-det}). \\
$T_{*,\nu},\ \Sigma_\nu,\ I_{0,\nu}$ & Linewidth-broadening temperature scale, cooling coefficient, and derived intensity scale entering the master curve of Sec.~\ref{sec:energy-balance}. \\
$\mathcal W_\nu,\ Q_{nm}$ & Cooling/onset factor (Eq.~\eqref{eq:cooling-factor}) and linear--nonlinear closure (Eq.~\eqref{eq:Qnm-def}); $Q_{nm}=1$ under shared cooling and common $\beta$. \\
\end{tabular}
\end{ruledtabular}
\end{table*}

\paragraph*{Summary of results.}
Our main goal is to separate universal dielectric physics from device-specific near-field coupling in a form that maps directly onto measured peak amplitudes~\cite{Bandurin2022NatPhys,Yahniuk2025}. The central outcomes are threefold. First, because the Bernstein dispersion is locally flat near each turning point~\cite{Roldan2011PRB}, the $q$-integral concentrates and the BM peak absorption factorizes into launcher, splitting $\Omega_n$ (which encodes both the Coulomb coupling and the $J_n^2$ dependence on cyclotron harmonic)~\cite{Kapralov2022PRB}, turning-point enhancement, and an intrinsic dielectric residue $\mathcal G_n^{\rm int}$. Second, at fixed excitation frequency all BM harmonics probe the same in-plane wavevector $q\approx\omega/v_F$ at leading order, so launcher factors cancel from inter-harmonic BM ratios. This leaves a parameter-free $m/n$ baseline and a single extracted reduced ratio $\mathcal R^{\rm res}_{nm}$ per harmonic pair, which measures the physically interesting departure from it. Third, the same factorized linear absorption feeds a minimal hot-electron model~\cite{BistrMac2009,SongReizer2012} that accounts for the distinct saturation behavior of BM and ordinary CR.

\medskip

Throughout, our emphasis is on the quasiclassical near-field regime relevant to the existing THz experiments; Landau-quantization corrections beyond that regime are left for future work. The paper is organized so that the main text develops the physical consequences in a reader-facing way, while the technical summation formulas and turning-point asymptotics are collected in Appendix~\ref{app:bessel}. Section~\ref{sec:Lambda} derives the factorized absorption coefficient and explains the meaning of each factor. Section~\ref{sec:ratios} derives the same-wavevector simplification, the three-factor decomposition, and (as a corollary) the polarization-collapse prediction for the launcher spectrum. Section~\ref{sec:protocol} translates those results into a concrete three-harmonic experimental protocol, while Section~\ref{sec:nonlinear} connects the linear absorption to nonlinear saturation by deriving the linewidth exponents from lineshape, separating absorption from detection, and establishing the inter-harmonic closure relation $Q_{nm}=\mathcal S_{nm}\mathcal R^{\rm res}_{nm}$. For orientation, Fig.~\ref{fig:bm-schematic} gives the geometry, Table~\ref{tab:notation} gives the notation, and Eq.~\eqref{eq:Lambda} gives the factorization.

\section{Absorption coefficient at each harmonic}\label{sec:Lambda}

Figure~\ref{fig:bm-schematic} and Table~\ref{tab:notation} already suggest the structure of the calculation. To connect that physical picture to an experimentally fitted resonance amplitude, it is useful to separate three factors: the launcher, the electronic loss function, and the turning-point projection. The metallic contact launches a continuum of in-plane momenta $q$; the graphene sheet absorbs those components according to its longitudinal loss function; and the BM peak emerges because the loss function becomes sharply concentrated near a turning point of the dispersion. In this language, for a long straight contact edge, translational invariance along the edge reduces the full $\int d^2\mathbf{q}$ to an integral over the dominant perpendicular wavevector component $q$, and the absorbed power at harmonic $n$ is~\cite{Bandurin2022NatPhys}
\begin{equation}
P_n(\omega) \propto \int dq\, |D(q,\phi)|^2\, \big[-\Im\,\varepsilon^{-1}(q,\omega)\big],
\label{eq:Pn}
\end{equation}
where $\varepsilon(q,\omega)$ is the longitudinal dielectric function including the gate-screened Coulomb interaction $V(q)$. The factor $-\Im\,\varepsilon^{-1}(q,\omega)$ is the usual \emph{loss function}: it measures how efficiently a longitudinal electric field of wavevector $q$ and frequency $\omega$ deposits energy into the electron system. The \emph{launcher spectrum} $|D(q,\phi)|^2$ encodes the spatial Fourier content of the evanescent near field produced by the metallic contacts~\cite{Bandurin2022NatPhys}, which act as antennas converting far-field radiation at frequency $\omega$ into finite-wavevector excitations in the 2DEG; it depends on both the contact geometry and the incident polarization angle $\phi$. The integral over $q$ is therefore essential rather than formal: a near-field experiment launches a \emph{band} of momenta, and the BM resonance selects a narrow part of that band. The launcher controls which momenta are available, the loss function determines which are absorbed, and the turning point concentrates the signal into a narrow part of the launched band.

Near the $n$th cyclotron overtone, the dielectric function takes the resonant form (Appendix~\ref{app:bessel})
\begin{equation}
\varepsilon_L(q,\omega)\simeq 1-\frac{\Omega_n(q)}{\Delta_\omega+i\Gamma_n},
\label{eq:epsL}
\end{equation}
where $\Delta_\omega = \omega - n\omega_c$ is the frequency detuning and $\Gamma_n$ is the fitted BM linewidth in frequency units (in the simplest model $\Gamma_n=\tau_p^{-1}$, but it is kept harmonic-dependent in the extraction protocol). The splitting $\Omega_n$ is
\begin{equation}
\Omega_n(q) = \frac{2\,\omega_p^2(q)\,n\omega_c}{\omega^2}\,h_n(qR_c),
\qquad
h_n(\zeta)\equiv\Bigl(\frac{n}{\zeta}\Bigr)^{\!2}J_n^2(\zeta),
\label{eq:Omega}
\end{equation}
with $\omega_p^2(q) \equiv n_0 q^2 V(q)/m_c$ the (squared) 2D plasma frequency at wavevector $q$~\cite{Hwang2007PRB,Wunsch2006NJP}, $V(q)$ the gate-screened Coulomb kernel (Appendix~\ref{app:bessel}), $R_c = v_F/\omega_c$ the cyclotron radius, and $v_F$ the Fermi velocity~\cite{CastroNeto2009RMP}.
Equation~\eqref{eq:epsL} is derived from the quasiclassical nonlocal longitudinal conductivity obtained from the Boltzmann equation~\cite{Kapralov2022PRB}. The factor $(n/\zeta)^2$ in $h_n$ arises from the kinematic relation between current and density through the continuity equation: it makes the pole residue of $\Omega_n$ proportional to $n\omega_c\,J_n^2(\zeta)$, which at fixed $\omega$ (where $n\omega_c=\omega$) is $n$-independent after factoring out the common prefactor $2\omega_p^2\omega/\omega^2$.

Here $J_n$ is the integer-order Bessel function of the first kind~\cite{DLMF}. For readers less familiar with why a Bessel function appears here, the origin is direct: in the semiclassical description~\cite{Dmitriev2012RMP} the phase factor associated with cyclotron motion is expanded in angular harmonics, and the corresponding Fourier coefficients are Bessel functions~\cite{Newberger1982}. The factor $J_n(qR_c)$ therefore measures how strongly a perturbation of wavevector $q$ couples to the $n$th cyclotron harmonic. The weight $h_n(qR_c)$ peaks near the turning point $qR_c\simeq n$, where the Bernstein dispersion flattens~\cite{Dmitriev2012RMP,Roldan2011PRB}.

\paragraph*{Three momentum scales.}
It is useful to distinguish three wavevectors that coincide only at leading order.
\begin{list}{}{\setlength{\labelwidth}{2.0em}\setlength{\leftmargin}{2.2em}\setlength{\labelsep}{0.2em}\setlength{\itemsep}{2pt}}
\item[$q_n$\;] $=n/R_c$: leading Airy turning-point momentum~\cite{DLMF,Watson1944}. Controls the fixed-$\omega$ cancellation: at resonance, $q_n=\omega/v_F\equiv k_\omega$ for every~$n$.
\item[$q_n^{(h)}$\;] Exact maximum of $h_n(qR_c)$. Sets the screening-independent baseline $\mathcal B^{(0)}_{nm}\simeq m/n$.
\item[$q_n^*$\;] Physical maximum of the full splitting $\Omega_n(q)$~\cite{Kapralov2022PRB,Roldan2011PRB}. Where the $q$-integral actually concentrates; evaluate $D$, $V$, $\mathcal K_n$ here. Screening-dependent through $\omega_p^2(q)$~\cite{Hwang2007PRB,Wunsch2006NJP}.
\end{list}
\noindent The cancellation is controlled by $q_n$; the baseline by $q_n^{(h)}$; the measured peak by $q_n^*$. All three agree at leading order, $q_n,q_n^{(h)},q_n^*\to k_\omega$ at fixed $\omega$; they differ at the $\sim\!20$--$50\%$ level in \emph{absolute} terms but only by a few percent \emph{pairwise} (Table~\ref{tab:qshift}). To avoid bookkeeping ambiguity, we write the main factorization directly at $q_n^*$. The scale $q_n=n/R_c$ used below is the leading Airy/Bessel turning scale of the $n$th harmonic residue. In a full device calculation the observed maximum is also shaped by the plasmon--harmonic crossing and by the launched $q$-spectrum. Departures from the narrow Airy window are not ignored; they are encoded in the physical $q_n^*$, the geometric correction $\mathcal C^{\rm geom}_{nm}$, and, ultimately, in the full-$q$ synthetic tests.

\paragraph*{Factorized peak absorption.}
Because the dispersion is locally parabolic about $q_n^*$, with $\Omega_n(q)\simeq \Omega_n(q_n^*)-\tfrac12 \mathcal K_n(q-q_n^*)^2$, the $q$-integral of Eq.~\eqref{eq:Pn} concentrates in a narrow window $\Delta q_n\sim(\Gamma_n/\mathcal K_n)^{1/2}$, and slowly varying prefactors can be evaluated on shell. The peak absorption at resonance then takes the factorized form (Appendix~\ref{app:bessel})
\begin{equation}
\Lambda_n(\phi,\omega) \propto |D(q_n^*,\phi)|^2\;\frac{\Omega_n(q_n^*)}{\sqrt{\mathcal{K}_n\,\Gamma_n}}\;\mathcal{G}_n^{\rm int},
\label{eq:Lambda}
\end{equation}
where $\mathcal{K}_n\equiv |\partial_q^2\Omega_n|_{q_n^*}$, $\Gamma_n$ is the linewidth in \emph{frequency} units, and $\mathcal{G}_n^{\rm int}$ is the intrinsic on-shell reduced peak factor defined below. Each factor has a distinct physical meaning:
\begin{itemize}
\item $|D(q_n^*,\phi)|^2$: the near-field launcher, \emph{evaluated at the physical turning point}; set by contact geometry and polarization.
\item $\Omega_n(q_n^*)$: the splitting at the physical turning point, which encodes both the Coulomb coupling $\omega_p^2(q_n^*)$ and the cyclotron-harmonic form factor $h_n(\zeta_n^*)$.
\item $(\mathcal{K}_n\Gamma_n)^{-1/2}$: the turning-point enhancement from the flat dispersion, producing the square-root lineshape of Ref.~\cite{Bandurin2022NatPhys}.
\item $\mathcal{G}_n^{\rm int}$: the residual smooth factor of the loss function at the peak, defined below and interpreted as an intrinsic dielectric residue.
\end{itemize}
The mechanism is the turning-point structure of $\Omega_n(q)$, controlled by the Airy turning region of $J_n$; after rescaling by $n^{1/3}$, the harmonic dependence is inherited from the universal Airy profile. This yields the leading scalings $J_n^2(n)\sim n^{-2/3}$ and $\mathcal K_n\sim n^{2/3}$ (Appendix~\ref{app:bessel}), giving the reference ratio $I_n/I_m\simeq m/n$ in the common-linewidth limit.

\paragraph*{The intrinsic residue $\mathcal G_n^{\rm int}$.}
The intrinsic residue $\mathcal G_n^{\rm int}$ is defined by the on-shell extraction
\begin{equation}
\mathcal G_n^{\rm int}
\equiv
\frac{\sqrt{\mathcal K_n\,\Gamma_n}}{\Omega_n(q_n^*)}
\left[\int_{\rm tp} dq\,\bigl(-\Im\,\varepsilon^{-1}(q,\omega)\bigr)\right]_{\rm peak},
\label{eq:Gn-def}
\end{equation}
where the integral is over the turning-point window with the smooth launcher factor held fixed, evaluated at the resonance maximum in $\omega$. By construction $\mathcal G_n^{\rm int}$ contains only the intrinsic dielectric response at $q_n^*$: the launcher and the Coulomb coupling are \emph{explicitly} displayed in Eq.~\eqref{eq:Lambda} through $|D(q_n^*)|^2$ and $\Omega_n(q_n^*)=(2\omega_p^2(q_n^*)n\omega_c/\omega^2)h_n(\zeta_n^*)$, and therefore do not belong to $\mathcal G_n^{\rm int}$. This is the conceptual advantage of writing Eq.~\eqref{eq:Lambda} at $q_n^*$ rather than at the asymptotic $q_n$: the ``smooth residual'' is a well-defined reduced peak factor. In particular, the device-dependent difference between $q_n^*$ and the common leading-order $k_\omega$ does \emph{not} sit inside $\mathcal G_n^{\rm int}$; it appears as a distinct and calculable pairwise factor $\mathcal C^{\rm geom}_{nm}$ introduced in Sec.~\ref{sec:ratios}.

\paragraph*{From $\Lambda_n$ to photoresistance.}
The photoresistance signal is $\Delta R_n = \beta_n\,P_{{\rm abs},n}\propto \beta_n\,I_{\rm inc}\,\Lambda_n(\phi,\omega)$, with the bolometric transfer coefficient $\beta_n$ weakly $n$-dependent~\cite{Bandurin2022NatPhys,Yahniuk2025}; at the low intensities of interest this is consistent with the linear-in-$I_{\rm inc}$ regime reported in those references. When extracting from photoresistance data, it is convenient to define the \emph{effective} residue
\begin{equation}
\mathcal G_n^{\rm eff}\equiv \beta_n\,\mathcal G_n^{\rm int},
\label{eq:Geff-def}
\end{equation}
so that the photoresistance amplitude takes the form $A_n\propto |D(q_n^*)|^2\,\Omega_n(q_n^*)(\mathcal K_n\Gamma_n)^{-1/2}\,\mathcal G_n^{\rm eff}$. When $\beta_n$ is common across BM harmonics at a given contact, the extracted $\mathcal G_n^{\rm eff}/\mathcal G_m^{\rm eff}$ reduces to $\mathcal G_n^{\rm int}/\mathcal G_m^{\rm int}$; otherwise, the $\beta$-ratio contaminates the dielectric interpretation by an explicit known factor. This separation distinguishes three content levels: a calculable baseline (below), a calculable pairwise launcher/screening correction (Sec.~\ref{sec:ratios}), and an extracted effective residue ratio.

Equation~\eqref{eq:Lambda} assumes that the launcher spectrum and other smooth prefactors vary slowly across the turning-point window $\Delta q_n\sim(\Gamma_n/\mathcal K_n)^{1/2}$; if $|D(q,\phi)|^2$ varies appreciably on that scale---e.g.\ a steep launcher tail or a structured narrowband near field---the factorization receives controlled derivative corrections, and the normalized angular patterns [Eq.~\eqref{eq:spectrometer-angular} below] need not remain strictly $n$-independent. Multi-harmonic measurements at different $n$, $\phi$, gate voltage, and contact geometry then probe different factors in Eq.~\eqref{eq:Lambda} independently. (Throughout, $I_{\rm inc}$ denotes incident intensity and $I_n$ the low-power photoresistance amplitude at harmonic $n$, proportional to $\beta_n\Lambda_n$; the two should not be confused.) This factorized peak formula is the organizing equation for the rest of the paper: later sections ask which pieces cancel in ratios, which remain device-specific, and which feed directly into the nonlinear problem.

\begin{figure}[!htbp]
\includegraphics[width=\columnwidth]{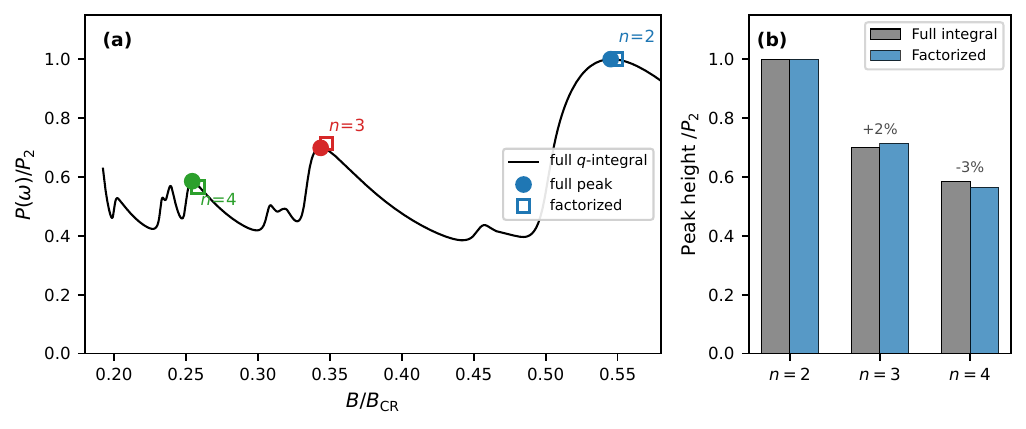}
\caption{Numerical benchmark of the factorized peak formula against the full $q$-integral~\eqref{eq:Pn} with the quasiclassical nonlocal dielectric function. (a)~Absorbed power vs.\ $B/B_{\rm CR}$ (normalized to the $n=2$ peak): filled circles mark full-integral peaks, open squares mark the factorized predictions (offset slightly in $B$ for visibility). (b)~Peak heights normalized to $n=2$: the factorized approximation agrees with the full integral to $+2\%$ ($n=3$) and $-3\%$ ($n=4$). Parameters: $f=2.54$~THz, $\alpha_g=0.51$, $k_F=3.1\times 10^8$~m$^{-1}$, $\Gamma/\omega_c=0.03$, unscreened Coulomb, model launcher~\eqref{eq:Dq}.}
\label{fig:benchmark}
\end{figure}

\paragraph*{Numerical validation.}
Figure~\ref{fig:benchmark} validates the factorized formula against the full $q$-integral~\eqref{eq:Pn} evaluated with the quasiclassical nonlocal dielectric function~\cite{Volkov2014PRB,Kapralov2022PRB} for realistic graphene parameters ($\alpha_g\approx 0.51$, $k_F\approx 3.1\times 10^8\,\mathrm{m}^{-1}$, $\Gamma/\omega_c=0.03$, $f=2.54\,\mathrm{THz}$, unscreened Coulomb, model launcher). The peak heights at $n=2,3,4$, normalized to $n=2$, agree to within $2$--$6\%$. More specifically, the inter-harmonic ratios $I_2/I_3=1.43$ (full) vs.\ $1.40$ (factorized) and $I_3/I_4=1.19$ (full) vs.\ $1.26$ (factorized) both lie close to the $m/n$ baseline ($1.50$ and $1.33$), confirming that the narrow-turning-window factorization captures the dominant ordering and that subleading corrections are at the few-percent level.

\paragraph*{Bessel building block.}
The on-shell splitting $\Omega_n(q_n)$ [Eq.~\eqref{eq:Omega}] is built from the kinematic weight $h_n(\zeta)=(n/\zeta)^2 J_n^2(\zeta)$, whose value at $\zeta=n$ reduces to $J_n^2(n)$. The uniform Debye--Olver asymptotics~\cite{DLMF,Watson1944} gives (Appendix~\ref{app:bessel})
\begin{equation}
J_n^2(n) \sim \frac{2^{2/3}\operatorname{Ai}^2(0)}{n^{2/3}} = \frac{0.2001}{n^{2/3}},
\label{eq:Jn2}
\end{equation}
where $\operatorname{Ai}$ denotes the Airy function~\cite{DLMF}, with $\operatorname{Ai}(0)=3^{-2/3}/\Gamma(2/3)\approx 0.3550$. The Airy function enters because a Bessel function near its turning point has a universal Airy profile; the exponent $2/3$ is the direct signature of that turning-point geometry. The approximation Eq.~\eqref{eq:Jn2} is already accurate to $1.2\%$ at $n=2$ and improves as $n^{-1}$ beyond (exact values $J_n^2(n)=0.12449,\,0.09552,\,0.07903,\,0.06819$ for $n=2,3,4,5$).
The decay is algebraic ($n^{-2/3}$), not exponential: higher harmonics remain observable with weakly structured launchers, provided the splitting remains larger than the linewidth. Physically, the same turning-point geometry that produces the square-root BM lineshape also produces this ordering: because the resonance is controlled by a universal fold of the dispersion, the peak heights fall only as $n^{-2/3}$ rather than being exponentially fragile, so it does not rely on fine-tuned screening or device geometry~\cite{DLMF}. However, $J_n^2(n)$ alone does not set the fixed-frequency inter-harmonic ratios; the curvature $\mathcal{K}_n\propto n^{2/3}$ (Sec.~\ref{sec:ratios}) also contributes, giving a net baseline $\mathcal{B}^{(0)}_{nm}\simeq m/n$ rather than the bare Bessel ratio. Both the $n^{-2/3}$ peak decay and the $n^{2/3}$ curvature growth originate in the same Bessel-to-Airy crossover at $\zeta\simeq n$ (Appendix~\ref{app:bessel}); the $m/n$ baseline is fixed entirely by that universal turning-point geometry.

\begin{figure}[!htbp]
\includegraphics[width=\columnwidth]{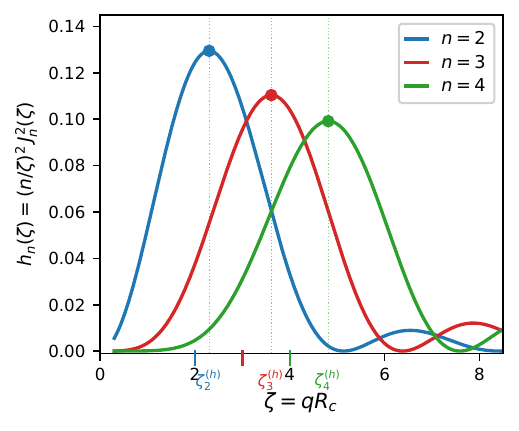}
\caption{The function $h_n(\zeta)=(n/\zeta)^2 J_n^2(\zeta)$ entering the splitting $\Omega_n$ [Eq.~\eqref{eq:Omega}], for $n=2,3,4$. Filled circles mark the Bessel peaks $\zeta_n^{(h)}$ (to be distinguished from the physical turning point $\zeta_n^*$ of the full $\Omega_n$, which is shifted outward by the $q$-slope of $\omega_p^2$; see Table~\ref{tab:qshift}); the $n^{-2/3}$ decay of the peak heights and the convergence of $\zeta_n^{(h)}/n$ toward a common value are both visible. The curvature at the peak controls $\mathcal{K}_n$ (Sec.~\ref{sec:ratios}), and the inter-harmonic peak shift controls the pairwise launcher correction (Sec.~\ref{sec:ratios}).}
\label{fig:hn}
\end{figure}

\section{Same-wavevector simplification at fixed frequency}\label{sec:ratios}

\paragraph*{Bernstein modes are not cyclotron resonance.}
Before analyzing inter-harmonic ratios, we note that the prominent peak near the fundamental ($n=1$) in experiment is \emph{not} a Bernstein mode: it is ordinary cyclotron resonance, a $q\approx 0$ far-field Lorentzian with $\propto\tau_p$ peak scaling and circular-polarization selectivity~\cite{Dmitriev2012RMP}, physically separate from the finite-$q$, near-field, turning-point-enhanced BM channel with square-root lineshape and $\propto\sqrt{\tau_p}$ scaling~\cite{Bandurin2022NatPhys}. The fixed-frequency BM inter-harmonic sequence discussed below is defined over $n\ge 2$; the $n=1$ peak is excluded. Launcher visibility is a separate issue from fixed-frequency ratios. A metallic contact of width $\ell$ produces a band-pass near-field spectrum peaked near $q_\star\sim 1/\ell$~\cite{Bandurin2022NatPhys}. If the common momentum $k_\omega=\omega/v_F$ lies in a launcher tail, all BM harmonics weaken together --- but this affects the overall BM signal, not the ratios among overtones. Those ratios depend on the intrinsic $n$-dependence derived next.

\paragraph*{Same-wavevector simplification.}
We now come to the central simplification of the paper. The principal simplification of Eq.~\eqref{eq:Lambda} emerges when comparing harmonics at fixed excitation frequency $\omega$, sweeping $B$ through the resonance fields $B_n = B_{\rm CR}/n$~\cite{Bandurin2022NatPhys,Yahniuk2025}. The key observation is that, at the bare resonance field $B_n$ for harmonic $n$ (i.e.\ where $n\omega_c=\omega$), the leading-order turning-point momentum takes a value that is independent of~$n$. Explicitly: at field $B_n$, the cyclotron frequency is $\omega_c(B_n)=\omega/n$, the cyclotron radius is $R_c(B_n)=n v_F/\omega$, and the leading-order turning-point momentum is
\begin{equation}
q_n^{(0)} = \frac{n}{R_c(B_n)} \simeq \frac{\omega}{v_F} \equiv k_\omega.
\label{eq:qfixed}
\end{equation}
This is independent of $n$. At fixed $\omega$, all BM harmonics probe the same leading-order launcher wavevector $k_\omega$. Every explicitly geometric factor in Eq.~\eqref{eq:Lambda} --- launcher spectrum, gate screening, and other smooth prefactors evaluated at $q_n$ --- is therefore sampled at the same momentum.

Because all device-dependent factors are evaluated at the same leading-order wavevector, they drop out when we take the ratio of two harmonics. The harmonic ratio reduces to
\begin{equation}
\frac{I_n}{I_m}\bigg|_{\omega\,\mathrm{fixed}}
= \frac{h_n(\zeta_n^{(h)})}{h_m(\zeta_m^{(h)})}
\sqrt{\frac{\mathcal{K}_m^{(0)}\,\Gamma_m}
{\mathcal{K}_n^{(0)}\,\Gamma_n}}\;
\frac{\mathcal{G}_n^{\rm eff}}{\mathcal{G}_m^{\rm eff}},
\label{eq:ratio-general}
\end{equation}
where $\zeta_n^{(h)}$ denotes the peak of $h_n(\zeta)=(n/\zeta)^2 J_n^2(\zeta)$, $\mathcal K_n^{(0)}\propto n^2|h_n''(\zeta_n^{(h)})|$ is the Bessel-peak curvature (Appendix~\ref{app:bessel}), and all device-specific factors entering through the common wavevector $k_\omega$ have cancelled. This is a leading-order baseline expression: the corrections from evaluating $h_n$ and $\mathcal K_n$ at the physical turning point $\zeta_n^*$ rather than the Bessel peak $\zeta_n^{(h)}$ are percent-level (Table~\ref{tab:qshift}) and are carried by the extracted effective residue defined below.

\paragraph*{Scope of the cancellation.}
Equation~\eqref{eq:ratio-general} is a leading-order statement about \emph{device} dependence: within the narrow-turning-window approximation~\cite{Bandurin2022NatPhys}, launcher spectrum~\cite{Bandurin2022NatPhys}, gate screening~\cite{Hwang2007PRB,Wunsch2006NJP}, and any other smooth factors that depend only on the common leading-order momentum $k_\omega$ cancel because they are evaluated at the same wavevector for every harmonic. With the bookkeeping of Sec.~\ref{sec:Lambda}, the ratio $\mathcal G_n^{\rm eff}/\mathcal G_m^{\rm eff}$ appearing in Eq.~\eqref{eq:ratio-general} is understood as a ratio of \emph{effective} residues $\mathcal G_n^{\rm eff}=\beta_n\mathcal G_n^{\rm int}$ carrying the bolometric transfer and the intrinsic dielectric residue only: the subleading device dependence from the $q_n^*-k_\omega$ shift is explicit, not hidden. We define the reduced ratio that extracts the non-baseline content from the data:
\begin{equation}
\mathcal R^{\rm res}_{nm}(\omega)
\equiv
\frac{I_n/I_m}{\mathcal B^{(0)}_{nm}\,\sqrt{\Gamma_m/\Gamma_n}},
\label{eq:residue-ratio}
\end{equation}
where $\mathcal B^{(0)}_{nm}$ is the parameter-free baseline defined below (Eq.~\eqref{eq:ratio-with-K}). This is the fixed-frequency observable that separates common device factors from the remaining response hierarchy. It is useful to display the decomposition explicitly:
\begin{equation}
\mathcal R^{\rm res}_{nm}
= \underbrace{\frac{|D(q_n^*)|^2}{|D(q_m^*)|^2}\,\frac{\omega_p^2(q_n^*)}{\omega_p^2(q_m^*)}}_{\displaystyle\mathcal C^{\rm geom}_{nm}}
\;\times\;\widetilde{\mathcal R}^{\rm eff}_{nm},
\label{eq:three-factor}
\end{equation}
where $\mathcal C^{\rm geom}_{nm}$ collects the pairwise \emph{launcher and screening} corrections at the physical turning points, and $\widetilde{\mathcal R}^{\rm eff}_{nm}$ is the extracted effective residue ratio that also absorbs the percent-level $h_n$/$\mathcal K_n$ shift corrections from $\zeta_n^{(h)}\to\zeta_n^*$:
\begin{align}
\widetilde{\mathcal R}^{\rm eff}_{nm}
&= \mathcal C_{nm}^{hK}\;\frac{\mathcal G_n^{\rm eff}}{\mathcal G_m^{\rm eff}},
\label{eq:Reff-tilde}\\[3pt]
\mathcal C_{nm}^{hK}
&\equiv
\frac{h_n(\zeta_n^*)/h_n(\zeta_n^{(h)})}{h_m(\zeta_m^*)/h_m(\zeta_m^{(h)})}
\nonumber\\[-2pt]
&\quad\times
\left[\frac{\mathcal K_m^*/\mathcal K_m^{(0)}}{\mathcal K_n^*/\mathcal K_n^{(0)}}\right]^{1/2}
= 1 + O(\text{few \%}).
\label{eq:ChK-def}
\end{align}
When $\mathcal C_{nm}^{hK}$ is neglected, $\widetilde{\mathcal R}^{\rm eff}_{nm}\approx \mathcal G_n^{\rm eff}/\mathcal G_m^{\rm eff}$. $\widetilde{\mathcal R}^{\rm eff}_{nm}$ is what remains once all calculable factors are divided out. The full inter-harmonic ratio thus takes the three-factor form
\begin{equation}
\frac{I_n}{I_m}\bigg|_{\omega\,\mathrm{fixed}}
= \mathcal B^{(0)}_{nm}\;\times\;\mathcal C^{\rm geom}_{nm}\;\times\;\widetilde{\mathcal R}^{\rm eff}_{nm},
\label{eq:three-factor-full}
\end{equation}
where each factor has a distinct status: $\mathcal B^{(0)}_{nm}\simeq m/n$ requires no free parameters; $\mathcal C^{\rm geom}_{nm}$ is calculable once the device geometry ($d$, $\ell$) is specified; and $\widetilde{\mathcal R}^{\rm eff}_{nm}$ is extracted from data and carries the remaining $n$-dependence after all calculable factors are removed. When harmonic-dependent linewidths are included, the full operational formula is
\begin{equation}
\frac{I_n}{I_m}\bigg|_{\omega\,\mathrm{fixed}}
= \mathcal B^{(0)}_{nm}\;\sqrt{\frac{\Gamma_m}{\Gamma_n}}\;\mathcal C^{\rm geom}_{nm}\;\widetilde{\mathcal R}^{\rm eff}_{nm}.
\label{eq:master-operational}
\end{equation}
The directly extractable reduced ratio is
\begin{equation}
\mathcal R^{\rm res}_{nm}
= \frac{I_n/I_m}{\mathcal B^{(0)}_{nm}\,\sqrt{\Gamma_m/\Gamma_n}}
= \mathcal C^{\rm geom}_{nm}\;\widetilde{\mathcal R}^{\rm eff}_{nm}.
\label{eq:Rres-operational}
\end{equation}
For reference, the ratio quantities are:
\begin{list}{}{\setlength{\labelwidth}{2.0em}\setlength{\leftmargin}{2.2em}\setlength{\labelsep}{0.2em}\setlength{\itemsep}{2pt}}
\item[$\mathcal B^{(0)}_{nm}$\;] Parameter-free fixed-frequency baseline, approximately $m/n$.
\item[$\sqrt{\Gamma_m/\Gamma_n}$\;] Linewidth factor multiplying the raw ratio $I_n/I_m$; divide by it when extracting residues.
\item[$\mathcal R^{\rm res}_{nm}$\;] Reduced ratio after removing the baseline and linewidth factor; includes any unresolved geometric correction.
\item[$\mathcal C^{\rm geom}_{nm}$\;] Optional launcher/screening correction from evaluating smooth factors at $q_n^*\neq q_m^*$.
\item[$\widetilde{\mathcal R}^{\rm eff}_{nm}$\;] Remaining effective residue ratio after all calculable factors are removed.
\end{list}

In practice, when a trusted launcher model is not available, the simpler two-factor form $I_n/I_m = \mathcal B^{(0)}_{nm}\sqrt{\Gamma_m/\Gamma_n}\,\mathcal R^{\rm res}_{nm}$ is preferred: the geometric correction is then implicitly folded into $\mathcal R^{\rm res}_{nm}$, and the three-factor decomposition~\eqref{eq:three-factor-full} becomes a refinement to be applied once the launcher is characterized. The synthetic recovery test below uses the full three-factor extraction, since the launcher generating model is specified.

When $\beta_n$ is common across BM harmonics at a given contact, $\widetilde{\mathcal R}^{\rm eff}_{nm}$ reduces to the pure intrinsic-dielectric residue ratio $\mathcal G_n^{\rm int}/\mathcal G_m^{\rm int}$; if the assumption fails~\cite{Bandurin2022NatPhys,Yahniuk2025}, the extracted $\widetilde{\mathcal R}^{\rm eff}_{nm}$ is $(\beta_n/\beta_m)\mathcal G_n^{\rm int}/\mathcal G_m^{\rm int}$, and the dielectric interpretation is correspondingly diluted by the known $\beta$-ratio.

\paragraph*{Extraction procedure.}
For one harmonic pair $(n,m)$, the linear extraction proceeds as follows.
\begin{enumerate}
\item Fit the low-power amplitudes $A_n,A_m$ and linewidths $\Gamma_n,\Gamma_m$.
\item Form the raw ratio $A_n/A_m$.
\item Divide by the baseline $\mathcal B^{(0)}_{nm}$ and, when linewidths are known, by $\sqrt{\Gamma_m/\Gamma_n}$. This gives $\mathcal R^{\rm res}_{nm}$.
\item If a trusted launcher/screening model is available, divide by $\mathcal C^{\rm geom}_{nm}$ to obtain $\widetilde{\mathcal R}^{\rm eff}_{nm}$.
\end{enumerate}
Test polarization collapse [Eq.~\eqref{eq:spectrometer-angular}] as an inter-harmonic check. If power sweeps are available, combine the linear and nonlinear extractions through $Q_{nm}=\mathcal S_{nm}\,\mathcal R^{\rm res}_{nm}$ [Eq.~\eqref{eq:Qnm-def}].

With three or more harmonics measured at the same frequency, the reduced ratios satisfy
\begin{equation}
\mathcal R^{\rm res}_{nm}(\omega)\;\mathcal R^{\rm res}_{mp}(\omega)
= \mathcal R^{\rm res}_{np}(\omega).
\label{eq:transitivity}
\end{equation}
Because $\mathcal R^{\rm res}_{nm}$ is by definition the ratio of two per-harmonic effective residues (up to the calculable $\mathcal C^{\rm geom}_{nm}$), Eq.~\eqref{eq:transitivity} is algebraically automatic. If all ratios are built from the same underlying amplitudes $(A_2,A_3,A_4)$, it holds \emph{exactly} even with noisy data; a nonzero failure therefore diagnoses extraction or model systematics (different background choices, pair-dependent fits, rounded baselines) rather than ordinary statistical noise. A genuine physics test is whether $\mathcal R^{\rm res}_{nm}(\omega)$ extracted from different devices or frequencies collapses onto a common smooth function, or whether the optional expansion~\eqref{eq:closure} below fits all measured harmonic pairs with one set of coefficients $g_i(\omega)$.

\paragraph*{Curvature scaling.}
To evaluate Eq.~\eqref{eq:ratio-general} we need to know how the dispersion curvature $\mathcal{K}_n$ depends on $n$ at fixed frequency. First set aside harmonic-dependent linewidths and take $\Gamma_n=\tau_p^{-1}$ in frequency units, independent of $n$. Then the universal part of the ratio is controlled by the curvature $\mathcal{K}_n$.

Linewidth anisotropy can be restored afterward by multiplying the raw ratio $I_n/I_m$ by $\sqrt{\Gamma_m/\Gamma_n}$~\cite{Moiseenko2025}. For long-range Coulomb scattering, $\Gamma_n$ is expected to increase with $n$, since forward-scattering-dominated disorder relaxes higher angular harmonics more efficiently. A representative anisotropy $\Gamma_3/\Gamma_2\approx 1.2$ gives $\sqrt{\Gamma_3/\Gamma_2}\approx 1.10$, an $\sim 10\%$ correction to $I_2/I_3$ comparable to $\mathcal C^{\rm geom}_{23}$. Equivalently, residue extraction divides the measured $I_2/I_3$ by this factor.

We now derive the curvature scaling. From Appendix~\ref{app:bessel}, the BM dispersion curvature at fixed $\omega$ is $\mathcal{K}_n \simeq |\partial_q^2\Omega_n|_{q_n^*}$~\cite{Roldan2011PRB,Kapralov2022PRB}, where $\Omega_n$ is the splitting~\eqref{eq:Omega}. At leading order we evaluate at the Bessel peak $\zeta_n^{(h)}$ (where $h_n'(\zeta_n^{(h)})=0$, so the omitted $\omega_p^2(q)$ cross term vanishes), and note that the remaining prefactor $2\omega_p^2 n\omega_c/\omega^2$ is $n$-independent at fixed $\omega$ (since $n\omega_c=\omega$):
\begin{equation}
\mathcal{K}_n \propto R_c^2\,\bigl|h_n''(\zeta_n^{(h)})\bigr|,
\qquad R_c = \frac{nv_F}{\omega}\propto n.
\label{eq:nscaling}
\end{equation}
In the Airy regime~\cite{DLMF}, $|h_n''|\propto n^{-4/3}$, giving $\mathcal{K}_n \propto n^2\times n^{-4/3} = n^{2/3}$. The curvature \emph{grows} with $n$: higher harmonics are reached at weaker magnetic fields (larger $R_c$), stretching the BM dispersion over a wider $q$ range and producing a sharper turning point. At the \emph{physical} turning point $\zeta_n^*>\zeta_n^{(h)}$ (shifted by the $q$-slope of $\omega_p^2$), $h_n'(\zeta_n^*)\neq 0$ and the cross-term $2A'(q_n^*)\,R_c\,h_n'(\zeta_n^*)$ in the full curvature $\partial_q^2\Omega_n = A''h_n + 2A'R_c h_n' + A R_c^2 h_n''$ contributes at the $\sim 8$--$37\%$ level (largest for $n=2$, declining with $n$), shifting $\mathcal K_n$ upward by $\sim 12$--$26\%$ relative to its $h_n$-peak value. Because the shift has weak $n$-dependence, the leading $n^{2/3}$ scaling survives; the numerical correction is absorbed into $\mathcal R^{\rm res}_{nm}$ rather than promoted to the explicit launcher/screening factor $\mathcal C^{\rm geom}_{nm}$.

\paragraph*{Inter-harmonic baseline.}
Equation~\eqref{eq:ratio-general} is the general fixed-frequency ratio, with the linewidth factor $\sqrt{\Gamma_m/\Gamma_n}$ explicit. It is useful to define the \emph{baseline}
\begin{equation}
\mathcal B^{(0)}_{nm}(\omega)\equiv \frac{h_n(\zeta_n^{(h)})}{h_m(\zeta_m^{(h)})}
\sqrt{\frac{m^2\,|h_m''(\zeta_m^{(h)})|}{n^2\,|h_n''(\zeta_n^{(h)})|}},
\label{eq:ratio-with-K}
\end{equation}
where $\zeta_n^{(h)}$ is the peak of $h_n(\zeta)=(n/\zeta)^2 J_n^2(\zeta)$ (Bessel peak, not the physical turning point $\zeta_n^*$ of $\Omega_n$). The baseline is the \emph{common-$\Gamma_n$, common-$\beta_n$} reduction of Eq.~\eqref{eq:ratio-general}: it is what remains of the universal $n$-dependence once harmonic-dependent linewidths and bolometric factors are set equal. Under those conventions the full fixed-frequency intensity ratio is
\begin{equation}
\frac{I_n}{I_m}\bigg|_{\omega\,\mathrm{fixed}}
= \mathcal B^{(0)}_{nm}\;\mathcal R^{\rm res}_{nm},
\label{eq:full-ratio}
\end{equation}
which separates the calculable baseline from the reduced ratio $\mathcal R^{\rm res}_{nm}=\mathcal C^{\rm geom}_{nm}\,\widetilde{\mathcal R}^{\rm eff}_{nm}$ [Eq.~\eqref{eq:three-factor}]. When $\Gamma_n$ varies with harmonic, a factor $\sqrt{\Gamma_m/\Gamma_n}$ multiplies Eq.~\eqref{eq:full-ratio}, typically a $10$--$30\%$ effect~\cite{Moiseenko2025}; when $\beta_n$ varies, the $\beta$-ratio multiplies $\widetilde{\mathcal R}^{\rm eff}_{nm}$. Numerically:
\begin{table}[!htbp]
\caption{baseline $\mathcal B^{(0)}_{nm}$ [Eq.~\eqref{eq:ratio-with-K}] for consecutive harmonics, computed from the $h_n$ peak alone and therefore screening-independent. The asymptotic scaling $m/n$ arises from $\Lambda_n\propto n^{-1}$ (Appendix~\ref{app:bessel}). The full ratio is $\mathcal B^{(0)}_{nm}\cdot\mathcal R^{\rm res}_{nm}$ [Eq.~\eqref{eq:full-ratio}]; the few-percent correction from the $\omega_p^2(q)$ slope is absorbed into $\mathcal R^{\rm res}_{nm}$ (see ``Scope'' paragraph).}
\label{tab:ratios}
\begin{ruledtabular}
\begin{tabular}{cccc}
$n/m$ & $\mathcal B^{(0)}_{nm}$ (exact) & $m/n$ & Deviation \\
\hline
2/3 & 1.54 & 1.50 & $+2.4\%$ \\
3/4 & 1.34 & 1.33 & $+0.7\%$ \\
4/5 & 1.25 & 1.25 & $+0.3\%$ \\
\end{tabular}
\end{ruledtabular}
\end{table}
The baseline is not near unity; it reflects the nontrivial $n$-dependence of both the splitting ($\Omega_n\propto n^{-2/3}$ from $h_n$) and the dispersion curvature ($\mathcal{K}_n\propto n^{2/3}$ from $R_c^2\propto n^2$). Equation~\eqref{eq:full-ratio} is therefore best read as a reference hierarchy: at fixed frequency, the parameter-free reference is $I_n/I_m=\mathcal B^{(0)}_{nm}\simeq m/n$. A measured departure from that value is not a failure of the peak formula but the signal of interest, namely $\mathcal R^{\rm res}_{nm}$ together with any quantified subleading corrections. The value of the decomposition is that division by $\mathcal B^{(0)}_{nm}$ makes the remainder interpretable.

\paragraph*{Robust content.}
The robust content of the fixed-frequency geometry is that overtones do \emph{not} scan different launcher momenta~\cite{Bandurin2022NatPhys}. Instead, they probe one common leading-order momentum $k_\omega=\omega/v_F$ and separate common device factors from the inter-harmonic ordering. After dividing out the calculable baseline $\mathcal B^{(0)}_{nm}\simeq m/n$, the reduced ratio~\eqref{eq:residue-ratio} isolates the smooth $n$-dependence that remains. A useful \emph{optional} fit ansatz is
\begin{equation}
\mathcal G_n^{\rm eff}(\omega) = \mathcal G^{\rm eff}(\omega)\left[1 + \frac{g_1(\omega)}{n} + \frac{g_2(\omega)}{n^2} + \cdots\right],
\label{eq:closure}
\end{equation}
under which
\begin{equation}
\widetilde{\mathcal R}^{\rm eff}_{nm}(\omega) = 1 + g_1(\omega)\!\left(\tfrac{1}{n}-\tfrac{1}{m}\right) + O(n^{-2},m^{-2}),
\label{eq:Reff-closure}
\end{equation}
and $\mathcal R^{\rm res}_{nm}(\omega)=\mathcal C^{\rm geom}_{nm}\,\widetilde{\mathcal R}^{\rm eff}_{nm}(\omega)$. When the bolometric transfer is common across harmonics, $\mathcal G_n^{\rm eff}\to\mathcal G_n^{\rm int}$, and Eq.~\eqref{eq:closure} is equivalently an ansatz on the intrinsic residue. The pure baseline $I_n/I_m = \mathcal B^{(0)}_{nm}$ corresponds to $\widetilde{\mathcal R}^{\rm eff}_{nm}=1$ and $\mathcal C^{\rm geom}_{nm}=1$. Whether that closure holds is a question to be settled by data, not assumed \textit{a priori}.

\paragraph*{Scope of the same-wavevector cancellation.}
The result $q_n=k_\omega$ refers to the leading-order turning point $qR_c=n$~\cite{Bandurin2022NatPhys,Roldan2011PRB}. Two distinct subleading $q$-shifts arise and must be distinguished.
The \emph{Bessel peak} of $h_n(\zeta)=(n/\zeta)^2 J_n^2(\zeta)$, which sets the $n$-scaling and baseline $\mathcal B^{(0)}_{nm}$, lies at $\zeta_n^{(h)}>n$~\cite{DLMF,Watson1944}.
The \emph{physical turning point} of the full splitting $\Omega_n(q)=(2\omega_p^2(q)/\omega)\,h_n(qR_c)$ is shifted further outward by the $q$-slope of $\omega_p^2(q)$~\cite{Hwang2007PRB,Wunsch2006NJP}; its location depends on the screening model (Table~\ref{tab:qshift}).
\begin{table}[!htbp]
\caption{Subleading momentum shifts at fixed frequency. $q_n^{(h)}$ is the Bessel-peak momentum of $h_n(\zeta)$; $q_n^*$ is the physical turning point of $\Omega_n(q)$, which additionally includes the $q$-slope of the screened Coulomb interaction. Both are normalized to the leading-order common momentum $k_\omega=\omega/v_F$. The ``deep-gated'' column corresponds to the strongly screened limit $k_\omega d\ll 1$, where $\omega_p^2(q)\propto q^2$; for finite gate distance the turning point lies between the unscreened and deep-gated values. The \emph{absolute} offset is $\sim\!20$--$50\%$, but the \emph{pairwise} mismatch between consecutive harmonics (which enters inter-harmonic ratios) is only a few percent.}
\label{tab:qshift}
\begin{ruledtabular}
\begin{tabular}{cccc}
$n$ & $q_n^{(h)}/k_\omega$ & $q_n^*/k_\omega$ (unscr.) & $q_n^*/k_\omega$ (deep-gated) \\
\hline
2 & 1.15 & 1.36 & 1.53 \\
3 & 1.20 & 1.31 & 1.40 \\
4 & 1.20 & 1.27 & 1.33 \\
5 & 1.19 & 1.24 & 1.28 \\
\end{tabular}
\end{ruledtabular}
\end{table}
The baseline $\mathcal B^{(0)}_{nm}\simeq m/n$ depends only on the Bessel peaks and is therefore screening-independent (Table~\ref{tab:ratios}). The geometric correction $\mathcal C^{\rm geom}_{nm}$, however, must be evaluated at the physical turning points $q_n^*$, which carry a screening-model dependence.

The \emph{absolute} offset ($q_n^*/k_\omega\approx 1.3$, common to all harmonics) cancels from inter-harmonic ratios; what matters is the \emph{pairwise} mismatch $|q_n^*-q_m^*|/k_\omega$. In other words, the large common shift away from $k_\omega$ does not spoil the ratio cancellation; only the small pairwise difference enters, and it is captured by $\mathcal C^{\rm geom}_{nm}$. For the unscreened case the pairwise $q$-shift is $\sim 5\%$ for $n=2/3$ and $\sim 4\%$ for $n=3/4$, declining slowly with $n$---unlike the $h_n$-peak mismatch, which is accidentally near-zero for $n=3/4$.

For the geometric correction we evaluate the model launcher~\eqref{eq:Dq}~\cite{Bandurin2022NatPhys} directly at the physical turning points $q_n^*$ (not via a linearized approximation or at the $h_n$ peaks). For a representative launcher with $k_\omega\ell= 1$ and $d/\ell= 0.75$ in the unscreened case:
\begin{equation}
\mathcal C^{\rm geom}_{nm}
= \frac{|D(q_n^*)|^2}{|D(q_m^*)|^2}\,\frac{\omega_p^2(q_n^*)}{\omega_p^2(q_m^*)}\approx
\begin{cases}
0.95 & n/m=2/3,\\
0.96 & n/m=3/4,\\
0.97 & n/m=4/5,
\end{cases}
\label{eq:Cgeom-values}
\end{equation}
i.e.\ pairwise corrections of $\sim 5\%$ for $n=2/3$ and $\sim 3$--$4\%$ for $n\ge 3/4$. The correction is largest for $n=2/3$ because the physical turning points $q_2^*$ and $q_3^*$ are the most widely separated (Table above). Across the parameter range $k_\omega\ell=0.8$--$1.2$, $d/\ell=0.5$--$1.0$, the $n=2/3$ correction varies from $\sim 2\%$ to $\sim 6\%$ (always $<1$: the launcher tail favors the higher harmonic, which has the smaller $q_m^*$, while the Coulomb factor $\omega_p^2\propto q$ favors the lower harmonic; for these parameters the launcher roll-off wins slightly). Since $\mathcal B^{(0)}_{23}=1.54$, even a $6\%$ geometric correction shifts the predicted full ratio only to $\sim 1.4$--$1.6$ before the effective residue --- the $m/n$ ordering still dominates.

\paragraph*{What is calculated and what is extracted.}
The three-factor form~\eqref{eq:three-factor-full} encodes this separation: $\mathcal B^{(0)}_{nm}\simeq m/n$ is parameter-free, $\mathcal C^{\rm geom}_{nm}$ is calculable from device geometry, and $\widetilde{\mathcal R}^{\rm eff}_{nm}$ is extracted from data; the lineshape exponents $s_{\rm BM}=1/2$ and $s_{\rm CR}=1$ and the angular-pattern $n$-independence [Eq.~\eqref{eq:spectrometer-angular} below] are likewise leading-order structural predictions. The assumptions behind the factorization and the corrections when each is relaxed are collected in Table~\ref{tab:validity} (App.~\ref{app:bessel}).

\paragraph*{Uncertainty propagation.}
Because the extracted observables are multiplicative, uncertainties are most directly propagated in log space (Appendix~\ref{app:uncertainty}). A nonzero failure of transitivity, $\mathcal R^{\rm res}_{23}\,\mathcal R^{\rm res}_{34}\neq\mathcal R^{\rm res}_{24}$, diagnoses extraction or model systematics rather than ordinary statistical noise when all ratios are built from the same underlying amplitudes. The dominant nuisances in a real photoresistance experiment --- harmonic-dependent linewidths at the $\sqrt{\Gamma_m/\Gamma_n}\sim 10$--$30\%$ level~\cite{Moiseenko2025}, the bolometric transfer $\beta_2/\beta_3$, and the shared-cooling assumption underlying $\mathcal S_{nm}$ --- are listed with their present status in Table~\ref{tab:validity} (App.~\ref{app:bessel}).

\paragraph*{Synthetic recovery test.}
To validate the extraction protocol on data against a known target, we generate synthetic BM peaks from the \emph{full} $q$-integral of Eq.~\eqref{eq:Pn}---not from the factorized approximation---using the resonant dielectric function Eq.~\eqref{eq:epsL} obtained from the Vavilov--Aleiner--Glazman-type kinetic equation~\cite{Kapralov2022PRB,Dmitriev2012RMP} with the gate-screened Coulomb kernel~\eqref{eq:Vq}~\cite{Hwang2007PRB,Wunsch2006NJP} and the launcher model~\eqref{eq:Dq}~\cite{Bandurin2022NatPhys}. Parameters are fixed as in the paper's illustrative example: $k_\omega\ell=1$, $d/\ell=0.75$, finite-gated Coulomb with $\omega_p^2(q)\propto q\,(1-e^{-2qd})$, and a plasmon strength yielding $\Omega_n^{\rm max}/\omega\sim 10^{-2}$ (deep turning-point regime). Harmonic-dependent linewidths $\Gamma_3/\Gamma_2=1.20$, $\Gamma_4/\Gamma_2=1.35$, motivated by the angular-harmonic-lifetime hierarchy of Ref.~\cite{Moiseenko2025}, provide a non-trivial target: $\widetilde{\mathcal R}^{\rm eff}_{nm}$ departs from unity at the $\sim 5$--$10\%$ level so that recovery is not automatically satisfied by $\widetilde{\mathcal R}^{\rm eff}_{nm}\equiv 1$. The computed physical turning points are $q_2^*/k_\omega=1.42$, $q_3^*/k_\omega=1.34$, $q_4^*/k_\omega=1.29$ (consistent with Table~\ref{tab:qshift} for the $d/\ell$ used).

\begin{figure}[!htbp]
\includegraphics[width=\columnwidth]{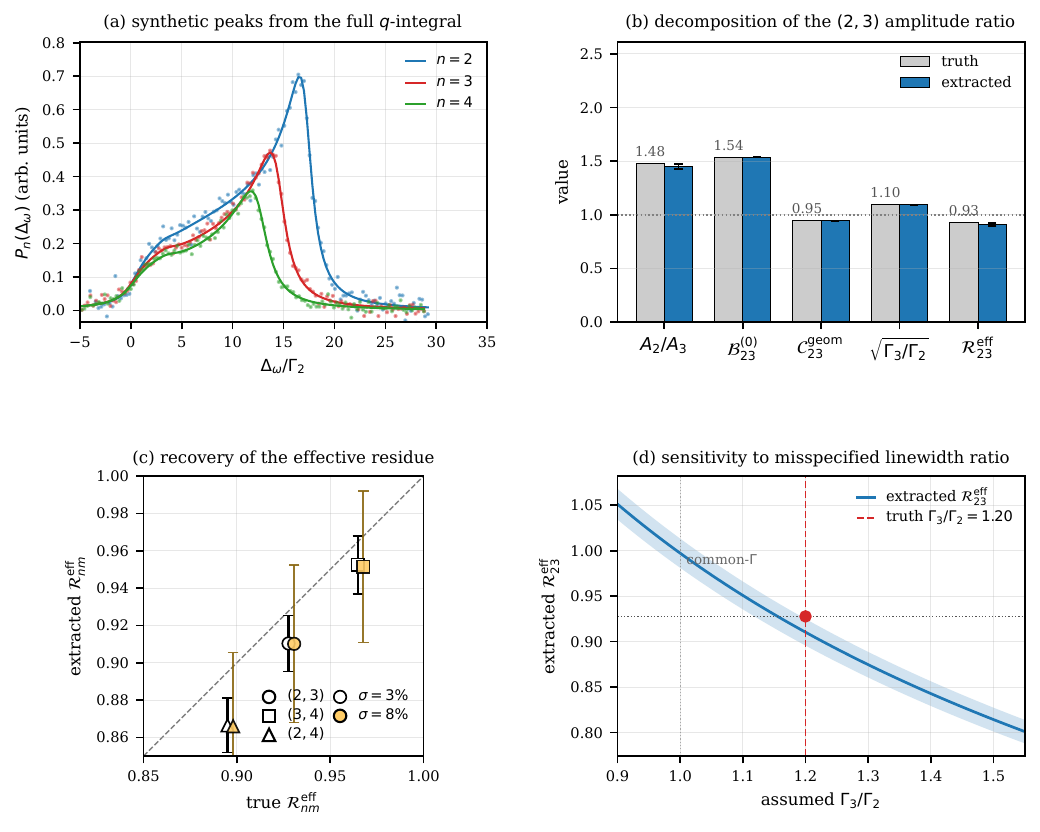}
\caption{Synthetic recovery test for the factorization protocol. (a)~Absorption profiles $P_n(\Delta_\omega)$ for $n=2,3,4$ generated by direct $q$-quadrature of Eq.~\eqref{eq:Pn} with the dielectric function Eq.~\eqref{eq:epsL}, the Coulomb kernel Eq.~\eqref{eq:Vq}, and the launcher Eq.~\eqref{eq:Dq}; solid lines, noiseless; dots, one 3\% per-point noise realization. (b)~Decomposition of the $(2,3)$ raw amplitude ratio $A_2/A_3$ (left bar of each pair: noiseless target from the profile; right bar: 400-realization median with $1\sigma$ Monte-Carlo error) into the baseline $\mathcal B^{(0)}_{23}=1.54$, the launcher/screening correction $\mathcal C^{\rm geom}_{23}=0.95$, the raw-ratio linewidth factor $\sqrt{\Gamma_3/\Gamma_2}=1.10$, and the effective residue $\widetilde{\mathcal R}^{\rm eff}_{23}$; equivalently, extraction of $\widetilde{\mathcal R}^{\rm eff}_{23}$ divides the measured ratio by $\sqrt{\Gamma_3/\Gamma_2}$. (c)~Extracted vs.\ true $\widetilde{\mathcal R}^{\rm eff}_{nm}$ for the three harmonic pairs at two noise levels ($3\%$, black; $8\%$, gold); the dashed line is $y=x$. (d)~Sensitivity of the extracted $\widetilde{\mathcal R}^{\rm eff}_{23}$ to a misspecified linewidth hierarchy: the curve shows the value an experimentalist would extract as a function of the \emph{assumed} ratio $\Gamma_3/\Gamma_2$; the true value is $1.20$ (red dashed), giving $\widetilde{\mathcal R}^{\rm eff}_{23}\approx 0.93$, while the common-$\Gamma$ assumption ($\Gamma_3/\Gamma_2=1$) would give $\approx 1.00$ (noiseless: $1.02$). The $\sim 10\%$ shift over the plausible range is the linewidth-knowledge-required part of the error budget. Shaded band: $16$--$84\%$ Monte-Carlo spread.}
\label{fig:recovery}
\end{figure}

From these inputs we compute $P_n(\Delta_\omega)$ by direct $q$-quadrature at $n=2,3,4$, add Gaussian noise per point, and fit a BM square-root lineshape~\cite{Bandurin2022NatPhys} to extract the peak amplitudes $A_n$. We then apply the extraction formula: divide by the precomputed baseline $\mathcal B^{(0)}_{nm}$ (Table~\ref{tab:ratios}), by the linewidth factor $\sqrt{\Gamma_m/\Gamma_n}$, and by the calculable geometric correction $\mathcal C^{\rm geom}_{nm}$ (Eq.~\eqref{eq:Cgeom-values}), leaving the effective residue $\widetilde{\mathcal R}^{\rm eff}_{nm}$. Repeating across 400 noise realizations at 3\% per-point noise yields the values shown in Fig.~\ref{fig:recovery}.

The extracted residues agree with the target values within the Monte-Carlo uncertainties:
\begin{equation}
\widetilde{\mathcal R}^{\rm eff}_{nm}\big|_{\rm extracted}
=
\begin{cases}
0.91\pm 0.03 & (n,m)=(2,3),\\
0.95\pm 0.03 & (n,m)=(3,4),\\
0.87\pm 0.03 & (n,m)=(2,4),
\end{cases}
\label{eq:Rint-synth}
\end{equation}
compared to the noiseless truths $0.93$, $0.96$, $0.89$ respectively. Because transitivity is algebraic when all three ratios are built from the same fitted amplitudes, we do not use it as a validation panel: it holds exactly by construction. Figure~\ref{fig:recovery}(d) instead quantifies the sensitivity of the extracted $\widetilde{\mathcal R}^{\rm eff}_{23}$ to an assumed linewidth hierarchy, showing that the common-$\Gamma$ assumption ($\Gamma_3/\Gamma_2=1$) shifts the extracted value by $\sim 10\%$ relative to the true $\Gamma_3/\Gamma_2=1.20$. Crucially, even in this exactly known setting the extracted $\widetilde{\mathcal R}^{\rm eff}_{nm}$ departs from unity by $5$--$13\%$: these deviations are the legitimate signature of the factorization-approximation residual and of the harmonic-dependent linewidth, not artifacts of the protocol. This is the feature that makes $\widetilde{\mathcal R}^{\rm eff}_{nm}$ a useful compressed observable.

A complementary misspecification stress test (Fig.~\ref{fig:adversarial}, Appendix~\ref{app:adversarial}) repeats the extraction under intentionally misspecified nuisance models. Over the tested ranges $\ell_{\rm ass}/\ell_{\rm true}=0.8$--$1.2$ and $d_{\rm ass}/\ell=0.5$--$1.0$ (i.e.\ $d_{\rm ass}/d_{\rm true}\approx 0.67$--$1.33$), the induced bias in the extracted $\widetilde{\mathcal R}^{\rm eff}_{23}$ remains below $\sim 2.3\%$ and $\sim 1.0\%$, respectively, while switching the extraction-side dielectric among unscreened, finite-gated, and deep-gated forms changes the result by at most $\sim 1.7\%$. By contrast, assuming a common linewidth when the generating model has $\Gamma_3/\Gamma_2=1.20$ produces a $\sim 9.5\%$ shift. Within these synthetic tests, moderate launcher/dielectric misspecification is therefore subleading to the linewidth hierarchy.

\paragraph*{Published-data comparison.}
A preliminary comparison with published power-sweep data~\cite{Yahniuk2025} is given in Appendix~\ref{app:published-data}; it illustrates a factor-of-$1.7$ ambiguity that the present protocol is designed to resolve. A purpose-built single-frequency multi-harmonic measurement would convert that ambiguity into an unambiguous residue extraction.

\paragraph*{Polarization corollary at fixed $\omega$.}
Because the splitting $\Omega_n$, curvature $\mathcal K_n$, and residue $\mathcal G_n^{\rm int}$ are scalar functions of $q$ and $\omega$, the \emph{only} factor in Eq.~\eqref{eq:Lambda} carrying the incident polarization angle $\phi$ is the launcher $|D(q,\phi)|^2$. At fixed $\omega$, all BM harmonics probe the common leading-order momentum $k_\omega$; their normalized angular patterns are therefore $n$-independent,
\begin{equation}
\frac{I_n(\phi)}{I_n(\phi_0)} = \frac{I_m(\phi)}{I_m(\phi_0)} \qquad (\text{all } n,m),
\label{eq:spectrometer-angular}
\end{equation}
whereas ordinary CR retains its circular-polarization selectivity~\cite{Dmitriev2012RMP}, although photoconductive CR in realistic 2DES structures can exhibit helicity immunity due to local near-field and disorder effects~\cite{Monch2022PRB} --- consistent with the helicity-insensitive but linearly-polarization-selective BM response of Ref.~\cite{Yahniuk2025}. Deviations from the $n$-independence would indicate that $|D(q,\phi)|^2$ varies appreciably over the turning-point window, or that the $q_n^*-k_\omega$ shift is resolved. When \emph{frequency} is varied, each $\omega$ selects a different effective sampling momentum $q_n^*(\omega)$ (not the bare $k_\omega$; see Table~\ref{tab:qshift}), so multi-frequency polarization sweeps map the launcher spectrum across momentum --- though three frequencies as in Refs.~\cite{Bandurin2022NatPhys,Yahniuk2025} are marginal; several are needed for a meaningful momentum-resolved profile.

\section{Experimental protocol: a three-harmonic ratio test}\label{sec:protocol}
The minimum data set is a single-frequency measurement of three BM harmonics, ideally $n=2,3,4$, with low-power amplitudes, linewidths, and polarization sweeps taken on the same device. Three harmonics are enough to extract two independent reduced ratios and to check the algebraic closure $\mathcal R^{\rm res}_{23}\mathcal R^{\rm res}_{34}=\mathcal R^{\rm res}_{24}$. At the carrier densities and THz frequencies used in Refs.~\cite{Bandurin2022NatPhys,Yahniuk2025} ($n_e\approx 2.8\times 10^{12}\,\mathrm{cm}^{-2}$ and $f=0.69$, $1.63$, $2.54$~THz), this is already enough to test the fixed-frequency baseline and the angular-collapse prediction. What three frequencies are \emph{not} enough for is full launcher tomography across momentum.

First, use the fixed-frequency ratios as a baseline extraction. Measure the low-power slopes $A_2,A_3,A_4$ and divide the raw ratios by the parameter-free baselines $\mathcal B^{(0)}_{23}=1.54$ and $\mathcal B^{(0)}_{34}=1.34$ (together with the linewidth factor $\sqrt{\Gamma_m/\Gamma_n}$ if the $\Gamma_n$ are independently known). If $A_2/A_3$ differs from $3/2$, the deviation is not a failure of the decomposition; it is $\mathcal R^{\rm res}_{23}$, the quantity carrying the actual physics. The pairwise geometric correction $\mathcal C^{\rm geom}_{nm}$ is only a few percent ($\sim 2$--$6\%$), whereas the dominant theoretical nuisance is harmonic-dependent linewidths, typically at the $\sim 10$--$20\%$ level and potentially $\sim 20$--$30\%$ for the $2/3$ pair if $\tau_2/\tau_3\sim 1.5$--$2$~\cite{Moiseenko2025}. The synthetic recovery test (Fig.~\ref{fig:recovery}) demonstrates that under this protocol the target effective residue is recovered within $\sim 3\%$ at realistic noise levels.

Second, use polarization to test the common-momentum prediction rather than to re-measure amplitudes. At one fixed frequency, record $A_n(\phi)$ for $n=2,3,4$ and check whether the normalized patterns $A_n(\phi)/A_n(\phi_0)$ collapse onto one curve [Eq.~\eqref{eq:spectrometer-angular}]; a linearly-polarization-selective, helicity-insensitive BM response of the kind reported in Ref.~\cite{Yahniuk2025} is a first indication that the launcher is the angular-pattern-carrying factor. This fixed-frequency test is sufficient with one frequency because it is an inter-harmonic test at common leading-order $k_\omega$, not a momentum scan. Frequency variation serves a different purpose, namely mapping the launcher spectrum~\cite{Bandurin2022NatPhys} across momentum. For that, the absolute momentum calibration is controlled by the physical turning point $q_n^*(\omega)$ rather than by the bare $k_\omega=\omega/v_F$; at the frequencies used in Refs.~\cite{Bandurin2022NatPhys,Yahniuk2025} the experiment samples only $k_\omega\approx 4.3$, $10.2$, $16.0$~$\mu\mathrm{m}^{-1}$, or $q_2^*\approx 6$--$24$~$\mu\mathrm{m}^{-1}$ depending on screening. Three such points are marginal for mapping a launcher band-pass; several well-separated frequencies would make the launcher measurement conclusive.

Third, if power-dependent data are available~\cite{Yahniuk2025}, test both the BM master curve and whether the cooling geometry~\cite{BistrMac2009,SongReizer2012} is shared across harmonics. From each raw sweep construct $a_n(I_{\rm inc})=\Delta R_n(I_{\rm inc})/(A_n I_{\rm inc})$ and collapse the traces using the corresponding $I_{0,n}$ or onset scale; BM traces from different harmonics should lie on the same BM master curve. Then form the single number $Q_{nm}=\mathcal S_{nm}\mathcal R^{\rm res}_{nm}$ [Eq.~\eqref{eq:Qnm-def}]. If $Q_{nm}=1$ within errors, the joint assumptions of shared cooling geometry ($\mathcal W_n=\mathcal W_m$) and common bolometric transfer ($\beta_n=\beta_m$) are self-consistent; if not, the deviation diagnoses harmonic-dependent cooling geometry, harmonic-dependent bolometric transfer, or a breakdown of the common-$q$ narrow-window approximation.
\section{Nonlinear saturation}\label{sec:nonlinear}

The preceding sections derived a factorized linear absorption and used it to build an inter-harmonic extraction protocol. The nonlinear part of the paper does not attempt to construct a complete microscopic theory of hot-electron cooling. Its purpose is to determine what power-dependent BM amplitudes must satisfy \emph{if the same factorized absorption coefficient that controls the low-power ratios also controls the onset of saturation}.

\subsection{Linewidth exponents from line shape}\label{sec:exponents}

The key structural input is how the peak absorption scales with the scattering-broadened linewidth~$\Gamma$.

Near a BM turning point, the dispersion is locally quadratic: $\omega-\omega_{\rm BM}(q)\simeq \Delta + \tfrac12 \mathcal K_n(q-q_n^*)^2$. At resonance ($\Delta=0$), the $q$-integrated peak therefore scales as
\begin{equation}
\alpha_{\rm BM}^{\rm peak}
\propto
\int dq\,\frac{\Gamma}
{\bigl[\tfrac12\mathcal K_n(q-q_n^*)^2\bigr]^2+\Gamma^2}
\propto (\mathcal K_n\Gamma)^{-1/2}.
\label{eq:sBM-derivation}
\end{equation}
For ordinary cyclotron resonance the peak is Lorentzian, giving $\alpha_{\rm CR}^{\rm peak}\propto\Gamma^{-1}$. Defining the linewidth exponent $s_\nu$ by $\alpha_\nu^{\rm peak}\propto\Gamma^{-s_\nu}$, this yields
\begin{equation}
s_{\rm BM}=\tfrac{1}{2},\qquad s_{\rm CR}=1.
\label{eq:exponents}
\end{equation}
This is a direct consequence of the same turning-point physics used in the linear factorization; no hot-electron model is needed. The same square-root turning-point scaling underlies the analytical BM plateau estimate in the Supplementary Information of Ref.~\cite{Bandurin2022NatPhys}, where the peak absorption scales as $\sqrt{\tau_p}$ in the simplest linewidth model. It also gives a robust experimental check independent of the cooling ansatz: for each resonance type,
\begin{equation}
\frac{\Delta R_\nu(I)}{I_{\rm inc}}
\left[\frac{\Gamma_\nu(I)}{\Gamma_{\nu,L}}\right]^{s_\nu}
\approx\text{constant},
\label{eq:lineshape-test}
\end{equation}
which tests the nonlinear mechanism at the lineshape level, not only through fitted saturation powers. Equation~\eqref{eq:lineshape-test} is an absorption-level test; in photoresistance data it holds either when $\beta_\nu$ is power-independent over the sweep or after correcting for the smooth $T_e$-dependence of the bolometric transfer using an independent resistance-versus-temperature calibration.

\subsection{Hot-electron energy balance}\label{sec:energy-balance}

\paragraph*{Scope of the model.}
We keep only the temperature dependence of the linewidth. Thermal corrections to the Drude weight, chemical potential, and screening are parametrically smaller in the degenerate regime ($k_BT_e\ll E_F$), while the launcher geometry is externally fixed. The first nonlinear correction to the peak height is therefore the linewidth broadening encoded in $\Gamma_\nu(T_e)$.

As a minimal phenomenological model, let the linewidth broaden with electron temperature as
\begin{equation}
\Gamma_\nu(T_e) = \Gamma_{\nu,L}\left[1 + \frac{T_e^2 - T_L^2}{T_{*,\nu}^2}\right],
\label{eq:Gamma-Te}
\end{equation}
where $\Gamma_{\nu,L} \equiv \Gamma_\nu(T_L)$ and $T_{*,\nu}$ is defined so that the linewidth doubles when $T_e^2 - T_L^2 = T_{*,\nu}^2$. The quadratic form is the simplest ansatz consistent with Fermi-liquid $T^2$ scattering~\cite{Dmitriev2012RMP,HwangHuDasSarma2007}; the qualitative conclusions below depend on Eq.~\eqref{eq:Gamma-Te} being monotonically increasing but not on the precise functional form. The peak absorption efficiency of a resonance of type $\nu$ then has the form
\begin{equation}
\alpha_\nu(T_e) = \alpha_{\nu,0}\left[\frac{\Gamma_{\nu,L}}{\Gamma_\nu(T_e)}\right]^{s_\nu},
\qquad s_{\rm BM} = \tfrac{1}{2},\quad s_{\rm CR} = 1,
\label{eq:alpha-Te}
\end{equation}
with $\alpha_{\nu,0} \equiv \alpha_\nu(T_L)$.

With acoustic-phonon cooling $P_{\rm cool} = \Sigma_\nu(T_e^\delta - T_L^\delta)$~\cite{BistrMac2009,SongReizer2012} (where $\delta = 4$ for clean graphene, $\delta = 3$ for disorder-assisted supercollision cooling), the steady-state energy balance becomes
\begin{equation}
I_{\rm inc}\,\alpha_{\nu,0}\left[1 + \frac{T_e^2 - T_L^2}{T_{*,\nu}^2}\right]^{-s_\nu}
= \Sigma_\nu\big(T_e^\delta - T_L^\delta\big).
\label{eq:master-balance}
\end{equation}
Here $\Sigma_\nu$ is channel-dependent: BM absorption heats a near-field hotspot localized near the metallic contact~\cite{Bandurin2022NatPhys,Yahniuk2025}, whereas CR heats the 2DEG more uniformly, so the effective cooling power per unit absorbed intensity generally differs between the two channels. The energy balance can be recast in dimensionless form by defining $X \equiv (T_e^2 - T_L^2)/T_{*,\nu}^2$, $r \equiv T_L/T_{*,\nu}$, and $I_{0,\nu} \equiv \Sigma_\nu T_{*,\nu}^\delta / \alpha_{\nu,0}$:
\begin{equation}
\frac{I}{I_{0,\nu}} = \Big[(r^2 + X)^{\delta/2} - r^\delta\Big](1+X)^{s_\nu}.
\label{eq:master-curve}
\end{equation}

\subsection{Absorption versus detection}\label{sec:abs-det}

For each BM harmonic, the low-power \emph{absorption} efficiency $\alpha_{n,0}^{\rm BM}$ is proportional to the factorized absorption coefficient $\Lambda_n$ of Eq.~\eqref{eq:Lambda}, with the proportionality constant set by the local field enhancement and active hot-spot area. The \emph{measured} low-power photoresistance slope is
\begin{equation}
A_n = \beta_n\,\alpha_{n,0}^{\rm BM},
\label{eq:An-beta}
\end{equation}
where $\beta_n$ is the bolometric/photoresistance transfer coefficient~\cite{Bandurin2022NatPhys,Yahniuk2025}. The energy balance~\eqref{eq:master-balance} depends on the \emph{absorption} coefficient $\alpha_{n,0}$, not on the detector transfer; $\beta_n$ enters only when connecting to the measured photoresistance. This distinction matters for the closure relation below, because $Q_{nm}=1$ requires not only shared cooling but also $\beta_n=\beta_m$.

More explicitly, in the notation of Sec.~\ref{sec:ratios}, the BM peak efficiency is
\begin{equation}
\alpha^{\rm BM}_{n,0}
\sim
\frac{|E_{\rm nf}/E_{\rm inc}|^2\,\mathcal{G}_n^{\rm int}\,|\widetilde D(q_n^*)|^2\,\Omega_n(q_n^*)}{\sqrt{\mathcal{K}_n\,\Gamma_{n,L}}},
\label{eq:alpha-BM-curvature}
\end{equation}
evaluated at the physical turning point $q_n^*$, where $\widetilde D$ is the normalized launcher spectral shape. The measured amplitude $A_n$ is $\beta_n$ times this quantity, so $\mathcal G_n^{\rm eff}=\beta_n\mathcal G_n^{\rm int}$ as defined in Sec.~\ref{sec:ratios}.

\subsection{BM inter-harmonic nonlinear closure}\label{sec:closure}

Nonlinearity sets in when $X\sim 1$, giving a crossover intensity
\begin{equation}
I_{\times,\nu}
=
\frac{2^{s_\nu}\,\mathcal W_\nu}{\alpha_{\nu,0}},
\label{eq:onset-general}
\end{equation}
where the cooling/onset factor
\begin{equation}
\mathcal W_\nu \equiv \Sigma_\nu T_{*,\nu}^{\delta}\Big[(r_\nu^2+1)^{\delta/2}-r_\nu^\delta\Big],
\qquad r_\nu \equiv T_L/T_{*,\nu},
\label{eq:cooling-factor}
\end{equation}
encodes the local cooling geometry and scattering environment at each resonance~\cite{BistrMac2009,SongReizer2012}.

From Eqs.~\eqref{eq:onset-general} and~\eqref{eq:An-beta},
\begin{equation}
A_n\,I_{\times,n} = 2^{1/2}\,\beta_n\,\mathcal W_n.
\label{eq:AnI-product}
\end{equation}
If the BM harmonics share the same hot spot ($\mathcal W_n=\mathcal W_m$) and the bolometric transfer is weakly harmonic-dependent ($\beta_n\approx\beta_m$),
\begin{equation}
A_n\,I_{\times,n}\approx\text{constant}.
\label{eq:AnI-const}
\end{equation}
This has the same structure as the empirical pattern $A_n\cdot I_{s,n}\approx\text{const}$ reported in Ref.~\cite{Yahniuk2025}, where $I_{s,n}$ is their saturation intensity corresponding to $I_{\times,n}$ in the present notation. In this product all absorption-side factors in Eq.~\eqref{eq:alpha-BM-curvature}---launcher, splitting, curvature, linewidth, and $\mathcal G_n^{\rm int}$---cancel; the measured product retains only the readout factor $\beta_n$ and the cooling/onset factor $\mathcal W_n$. Thus the $q$-shift corrections affect $A_n$ and $I_{\times,n}$ separately, but not their product.

The ratio version provides a direct bridge between weak-power and nonlinear data. At fixed frequency, $q_n^*\simeq q_m^*\simeq k_\omega$ at leading order, so the smooth launcher and Coulomb factors cancel to lowest order. Because $I_{\times,n}$ is inversely proportional to the low-power absorption coefficient, the onset analogue removes the same baseline multiplicatively:
\begin{align}
\mathcal S_{nm}(\omega)
&\equiv
\frac{I_{\times,n}}{I_{\times,m}}\;\mathcal B^{(0)}_{nm}\,\sqrt{\frac{\Gamma_{m,L}}{\Gamma_{n,L}}}
\nonumber\\[-2pt]
&= \frac{\mathcal W_n}{\mathcal W_m}\frac{\mathcal G_m^{\rm int}}{\mathcal G_n^{\rm int}}\,\big[\mathcal C^{\rm geom}_{nm}\,\mathcal C_{nm}^{hK}\big]^{-1}.
\label{eq:onset-reduced}
\end{align}
The relation to the linear reduced ratio is
\begin{equation}
\mathcal S_{nm}(\omega) = \frac{\mathcal W_n}{\mathcal W_m}\,\frac{\beta_n}{\beta_m}\,\big[\mathcal R^{\rm res}_{nm}(\omega)\big]^{-1}.
\label{eq:reciprocity}
\end{equation}
This makes the assumptions explicit. Measure low-power amplitudes $A_n,A_m$ and linewidths to obtain $\mathcal R^{\rm res}_{nm}$ from the linear data; measure onset intensities $I_{\times,n},I_{\times,m}$ to obtain $\mathcal S_{nm}$ from the nonlinear data; then form
\begin{equation}
Q_{nm}\equiv \mathcal S_{nm}\,\mathcal R^{\rm res}_{nm}.
\label{eq:Qnm-def}
\end{equation}
Under shared cooling geometry ($\mathcal W_n=\mathcal W_m$) and common bolometric transfer ($\beta_n=\beta_m$),
\begin{equation}
Q_{nm}=1.
\label{eq:Qnm-unity}
\end{equation}
Equivalently,
\begin{equation}
Q_{nm} = \frac{A_n\,I_{\times,n}}{A_m\,I_{\times,m}},
\label{eq:Qnm-AnI}
\end{equation}
so $Q_{nm}=1$ is the ratio version of the statement that $A_nI_{\times,n}$ is harmonic independent. All Bessel, curvature, and linewidth factors cancel in this product. A systematic deviation of $Q_{nm}$ from unity diagnoses either harmonic-dependent cooling geometry ($\mathcal W_n\neq\mathcal W_m$) or harmonic-dependent bolometric transfer ($\beta_n\neq\beta_m$). The pairwise geometric correction $\mathcal C^{\rm geom}_{nm}$ cancels from $Q_{nm}$ when the same reduced ratio $\mathcal R^{\rm res}_{nm}$ of Eq.~\eqref{eq:residue-ratio} is used; geometry re-enters only if $\mathcal S_{nm}$ and $\mathcal R^{\rm res}_{nm}$ are constructed with inconsistent conventions, or if the factorized peak formula itself breaks down beyond the narrow-turning-window approximation.

\subsection{BM versus CR saturation}\label{sec:bm-vs-cr}

The exponent difference~\eqref{eq:exponents} fixes the \emph{shape} of the normalized saturation curves and gives a robust distinction between BM and CR. The absolute BM/CR saturation-power ordering, however, also depends on the low-power absorption efficiency and on the effective heated/cooling geometry.

In the low-lattice-temperature limit $T_L\ll T_{*,\nu}$, the master curve~\eqref{eq:master-curve} can be written directly in terms of the normalized peak amplitude
$a_\nu(I_{\rm inc})\equiv \alpha_\nu(I_{\rm inc})/\alpha_{\nu,0}$.
Eliminating $X$ gives
\begin{align}
\frac{I}{I_{0,\rm BM}} &= \frac{(1-a_{\rm BM}^2)^{\delta/2}}{a_{\rm BM}^{\delta+1}},
\label{eq:collapse-bm}\\
\frac{I}{I_{0,\rm CR}} &= \frac{(1-a_{\rm CR})^{\delta/2}}{a_{\rm CR}^{\delta/2+1}},
\label{eq:collapse-cr}
\end{align}
with $I_{0,\nu}\equiv \Sigma_\nu T_{*,\nu}^{\delta}/\alpha_{\nu,0}$.
Operationally, an experimentalist can construct the collapse directly from raw power sweeps~\cite{Yahniuk2025}: the vertical normalization is
\[
a_\nu(I_{\rm inc})=\frac{\Delta R_\nu(I_{\rm inc})}{A_\nu I_{\rm inc}},
\]
and the horizontal normalization is the single scale $I_{0,\nu}$ (fitted per trace or inferred from the onset).

The most robust nonlinear predictions are therefore:
(i)~BM power sweeps collapse with $s=1/2$;
(ii)~CR power sweeps collapse with $s=1$;
(iii)~for BM harmonics sharing a hotspot, $A_nI_{\times,n}\approx\text{const}$ and $Q_{nm}\approx 1$.
\noindent Quantitative BM/CR saturation-power ratios are less robust. The ratio of BM to CR onset intensities,
\begin{equation}
\frac{I_{\times,\rm BM}}{I_{\times,\rm CR}}
= \frac{\Sigma_{\rm BM}}{\Sigma_{\rm CR}}\,
\left(\frac{T_{*,\rm BM}}{T_{*,\rm CR}}\right)^{\!\delta}
\frac{\alpha_{{\rm CR},0}}{\alpha_{{\rm BM},0}}\,2^{-1/2},
\label{eq:onset-ratio}
\end{equation}
contains the structural exponent $2^{-1/2}$ from $s_{\rm BM}-s_{\rm CR}=-1/2$, but all remaining factors are channel- and device-dependent. For deeper saturation, if $I_f$ is the intensity at which the peak efficiency has fallen to fraction $f$ of its low-power value, the ratio involves an additional amplification factor:
\begin{equation}
\frac{I_f^{\rm BM}}{I_f^{\rm CR}}
\simeq \frac{\Sigma_{\rm BM}}{\Sigma_{\rm CR}}\,
\left(\frac{T_{*,\rm BM}}{T_{*,\rm CR}}\right)^{\!\delta}
\frac{\alpha_{{\rm CR},0}}{\alpha_{{\rm BM},0}}
\left(\frac{1+f}{f}\right)^{\!\delta/2}.
\label{eq:If-ratio}
\end{equation}
For clean graphene ($\delta=4$) and half-maximum ($f=1/2$), the extra factor is~$9$; for supercollision cooling ($\delta=3$)~\cite{SongReizer2012} it is $\sim\!5.2$. Because this coefficient depends sensitively on both $\delta$ and the precise value of $s_{\rm BM}$ (a $\pm 0.1$ shift in $s$ produces a factor-of-$4$ change), it should be regarded as structural rather than quantitatively predictive without intensity-dependent lineshape data.

\begin{figure}[!htbp]
\includegraphics[width=\columnwidth]{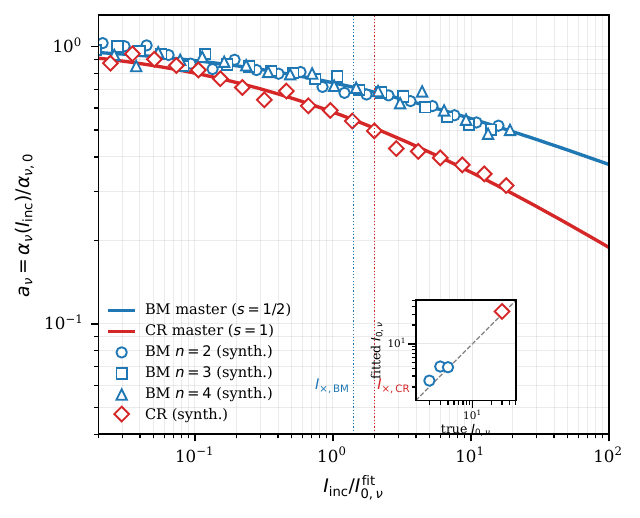}
\caption{Normalized saturation master curves in the low-$T_L$ limit for clean graphene ($\delta=4$), with mock power-sweep traces generated from the energy-balance model overlaid (symbols). The plotted quantity is the normalized peak amplitude $a_\nu=\alpha_\nu(I_{\rm inc})/\alpha_{\nu,0}=\Delta R_\nu(I_{\rm inc})/[A_\nu I_{\rm inc}]$ versus the scaled intensity $I_{\rm inc}/I_{0,\nu}^{\rm fit}$, where $I_{0,\nu}^{\rm fit}$ is fitted per trace as the one-parameter horizontal rescaling. Mock traces are generated by solving the energy balance Eq.~\eqref{eq:master-balance} for $X(I_{\rm inc})$ with $r=T_L/T_{*,\nu}=0.2$, adding $4\%$ multiplicative noise, and fitting $A_\nu$ on the lowest-power linear portion. BM traces at $n=2,3,4$ (open circles, squares, triangles) share a common $s_{\rm BM}=1/2$ and collapse onto the same BM master curve despite different $I_{0,n}$; CR (open diamonds, $s_{\rm CR}=1$) falls on its own curve. Dotted verticals mark onset intensities. Inset: fitted vs.\ true $I_{0,\nu}$ for each mock trace; the fit recovers the generating values with scatter of order $4$--$40\%$ (the wider scatter arising when the sweep does not extend deep into saturation).}
\label{fig:master-curves}
\end{figure}

\subsection{High-temperature behavior and magnetic-field scaling}\label{sec:high-T}

Reference~\cite{Yahniuk2025} reports that at elevated base temperature ($\sim 20$~K) the photoresponse remains linear over the available intensity range. Within the present model this is natural: for large $r_\nu=T_L/T_{*,\nu}$,
\begin{equation}
\mathcal W_\nu \simeq \tfrac{\delta}{2}\,\Sigma_\nu\,T_{*,\nu}^2\,T_L^{\delta-2},
\label{eq:W-highT}
\end{equation}
so $I_{\times,\nu}\propto T_L^{\delta-2}$ increases strongly with lattice temperature and the onset is pushed beyond the accessible range.

Within a single resonance type (e.g.\ comparing BM harmonics at different $B$), the structurally robust prediction is the $\sqrt{\tau_p}$ dependence of the BM peak height~\cite{Bandurin2022NatPhys}, which follows from the $\Gamma^{-1/2}$ turning-point scaling. A universal $B^{1/2}$ law for the BM-to-CR onset ratio is \emph{not} expected, because it additionally requires knowledge of the channel-dependent cooling ratio $\Sigma_{\rm BM}/\Sigma_{\rm CR}$ and of the field dependence entering through $q_n$, $|D(q_n)|^2$, $\Omega_n(q_n)$, and $\mathcal{K}_n$.

\paragraph*{Assumptions and observables.}
Table~\ref{tab:nonlinear-summary} summarizes the status of the nonlinear quantities and how each is experimentally accessed.

\begin{table}[!htbp]
\caption{Nonlinear quantities: status and experimental access.}\label{tab:nonlinear-summary}
\begin{tabular}{p{0.30\columnwidth}p{0.30\columnwidth}p{0.30\columnwidth}}
\hline\hline
Quantity & Status & Access \\
\hline
$s_{\rm BM}=1/2$, $s_{\rm CR}=1$ & Robust & Lineshape; Eq.~\eqref{eq:lineshape-test} \\[2pt]
$A_nI_{\times,n}\approx\text{const}$ & Shared hotspot $+$ $\beta_n\approx\beta_m$ & Power sweeps \\[2pt]
$Q_{nm}=1$ & Shared cooling $+$ common $\beta$ & Linear $+$ nonlinear data \\[2pt]
$\delta=3,4$ & Material/disorder dependent & Cooling-power fits \\[2pt]
$\mathcal W_\nu$ & Device/hotspot dependent & From $I_{\times,\nu}$ via Eq.~\eqref{eq:onset-general} \\[2pt]
Factor of $9$ at half-max & Ansatz-dependent ($\delta$, $s_{\rm BM}$) & Intensity-dependent lineshape \\[2pt]
\hline\hline
\end{tabular}
\end{table}
\FloatBarrier
\section{Conclusions and outlook}

\emph{Linear-response result.}\;
The peak absorption at each Bernstein harmonic factorizes, in the quasiclassical narrow-turning-window approximation, into the launcher spectrum at the physical turning point $q_n^*$, the splitting $\Omega_n(q_n^*)$, a turning-point enhancement $(\mathcal K_n\Gamma_n)^{-1/2}$, and an intrinsic dielectric residue $\mathcal G_n^{\rm int}$ (Sec.~\ref{sec:Lambda}). At fixed excitation frequency all BM overtones probe the same leading-order momentum $k_\omega=\omega/v_F$, so smooth device-dependent factors cancel at leading order from inter-harmonic ratios, leaving a parameter-free baseline $\mathcal B^{(0)}_{nm}\simeq m/n$ and an extracted reduced ratio $\mathcal R^{\rm res}_{nm}$ per harmonic pair; when a trusted launcher model is available, $\mathcal R^{\rm res}_{nm}$ further decomposes into a calculable geometric correction $\mathcal C^{\rm geom}_{nm}$ of a few percent and an effective residue $\widetilde{\mathcal R}^{\rm eff}_{nm}$. A controlled synthetic recovery test, generating peaks from the full $q$-integral with realistic noise and harmonic-dependent linewidths, recovers the target effective residue within Monte-Carlo errors (Fig.~\ref{fig:recovery}); a misspecification stress test (Fig.~\ref{fig:adversarial}) confirms that, within the tested smooth-launcher synthetic ranges, moderate launcher/dielectric misspecification induces only $\sim\!1$--$2\%$ shifts, with a common-$\Gamma$ assumption dominating the error budget at $\sim\!10\%$.

\emph{Experimental test.}\;
The three-harmonic single-frequency protocol of Sec.~\ref{sec:protocol}: low-power slopes $A_{2,3,4}$ plus polarization sweeps plus --- if available --- power sweeps, on one device. Two independent reduced ratios can be extracted, the algebraic closure checked, and the polarization-collapse prediction [Eq.~\eqref{eq:spectrometer-angular}] tested at leading-order common $k_\omega$. This is substantially more informative per data point than the multi-frequency launcher-tomography problem, which requires several well-separated frequencies. A practical consequence of the protocol is that it separates the linewidth factor $\Gamma_n^{-1/2}$ from the reduced amplitude residue, so that amplitude hierarchies are not confused with harmonic-dependent damping; this distinction is essential before interpreting high-order resonances in terms of electron-electron angular relaxation~\cite{Moiseenko2025}.

\emph{Nonlinear extension.}\;
The turning-point lineshape gives $s_{\rm BM}=1/2$ versus $s_{\rm CR}=1$ (Sec.~\ref{sec:exponents}), which is a direct consequence of the same factorization and requires no hot-electron model. For BM harmonics sharing a cooling geometry and bolometric transfer, the closure relation $A_nI_{\times,n}\approx\text{const}$ and $Q_{nm}=\mathcal S_{nm}\mathcal R^{\rm res}_{nm}=1$ links linear and nonlinear data through a single number (Sec.~\ref{sec:closure}). BM and CR power sweeps fall on distinct normalized master curves; quantitative saturation-power ratios (e.g.\ the factor of $9$ at half-maximum for $\delta=4$) depend on the cooling exponent and lineshape ansatz and should be regarded as structural rather than predictive (Table~\ref{tab:nonlinear-summary}).

\emph{Outlook.}\;
The same Bessel-product sum $S_{00}(\mu,\zeta)=\sum_n J_n^2(\zeta)/(n-\mu)$ that enters the present factorization also governs the transport analysis of Hall-field-induced resistance oscillations (HIRO)~\cite{Tierz2025}; its closed-form evaluation~\cite{Newberger1982} applies to both problems but at different poles and in different asymptotic regimes. In HIRO the relevant pole is imaginary, $\mu=i/\sqrt\chi$, set by the smooth-disorder correlation scale; the asymptotics are large-argument, $J_n^2(\zeta)\sim 2/(\pi\zeta)$. Here the pole is integer, $\mu=n$, set by the cyclotron harmonic; the asymptotics are turning-point, $J_n^2(n)\sim n^{-2/3}$ (App.~\ref{app:bessel}.7). Applying the same contour-integral closure to the unevaluated Bessel-product sums appearing in the kinetic model of Ref.~\cite{Kapralov2022PRB} could yield closed-form BM dispersions. Related factorization ideas apply to parabolic 2DEGs~\cite{Volkov2014PRB,Holland2004PRL}, bilayer graphene magnetoplasmon structures~\cite{Monch2023PRB}, and shear Bernstein modes in correlated 2D liquids~\cite{Afanasiev2023PRB}. Two extensions would strengthen the quantitative reach: (i)~Landau-quantization corrections~\cite{Goerbig2011RMP} beyond the quasiclassical regime, and (ii)~phase-resolved scattering-type scanning near-field optical microscopy (s-SNOM)~\cite{Fei2012Nature,Chen2012Nature,JiangShi2016NatMat} measurements accessing the coherent Airy oscillations underlying the projected square-root envelope.

\begin{acknowledgments}
This work was supported by the Shanghai Institute for Mathematics and Interdisciplinary Sciences (SIMIS-ID-2025-QT).
\end{acknowledgments}

\bibliographystyle{apsrev4-2}
\bibliography{bernstein_v34}

@article{Bernstein1958,
  author={I. B. Bernstein}, title={Waves in a Plasma in a Magnetic Field},
  journal={Phys. Rev.}, volume={109}, pages={10--21}, year={1958},
  doi={10.1103/PhysRev.109.10}}

@article{ChiuQuinn1974,
  author={K. W. Chiu and J. J. Quinn}, title={Plasma oscillations of a two-dimensional electron gas in a strong magnetic field},
  journal={Phys. Rev. B}, volume={9}, pages={4724--4732}, year={1974},
  doi={10.1103/PhysRevB.9.4724}}

@article{batke1985nonlocality,
  author={E. Batke and D. Heitmann and J. P. Kotthaus and K. Ploog}, title={Nonlocality in the Two-Dimensional Plasmon Dispersion},
  journal={Phys. Rev. Lett.}, volume={54}, pages={2367--2370}, year={1985},
  doi={10.1103/PhysRevLett.54.2367}}

@article{Lefebvre1998SST,
  author={J. Lefebvre and J. Beerens and Y. Feng and Z. Wasilewski and J. Beauvais and E. Lavall{\'e}e},
  title={{Bernstein} modes in a laterally modulated two-dimensional electron gas},
  journal={Semicond. Sci. Technol.}, volume={13}, pages={169--172}, year={1998},
  doi={10.1088/0268-1242/13/2/004}}

@article{Volkov2014PRB,
  author={V. A. Volkov and A. A. Zabolotnykh}, title={Bernstein modes and giant microwave response of a two-dimensional electron system},
  journal={Phys. Rev. B}, volume={89}, pages={121410}, year={2014},
  doi={10.1103/PhysRevB.89.121410}}

@article{Holland2004PRL,
  author={S. Holland and C. Heyn and D. Heitmann and E. Batke and R. Hey and K. J. Friedland and C.-M. Hu}, title={Quantized Dispersion of Two-Dimensional Magnetoplasmons Detected by Photoconductivity Spectroscopy},
  journal={Phys. Rev. Lett.}, volume={93}, pages={186804}, year={2004},
  doi={10.1103/PhysRevLett.93.186804}}

@article{Roldan2011PRB,
  author={R. Rold{\'a}n and M. O. Goerbig and J.-N. Fuchs}, title={Theory of {Bernstein} modes in graphene},
  journal={Phys. Rev. B}, volume={83}, pages={205406}, year={2011},
  doi={10.1103/PhysRevB.83.205406}}

@article{Kapralov2022PRB,
  author={K. Kapralov and D. Svintsov}, title={Ballistic-to-hydrodynamic transition and collective modes for two-dimensional electron systems in magnetic field},
  journal={Phys. Rev. B}, volume={106}, pages={115415}, year={2022},
  doi={10.1103/PhysRevB.106.115415}}

@article{CastroNeto2009RMP,
  author={A. H. Castro Neto and F. Guinea and N. M. R. Peres and K. S. Novoselov and A. K. Geim}, title={The electronic properties of graphene},
  journal={Rev. Mod. Phys.}, volume={81}, pages={109--162}, year={2009},
  doi={10.1103/RevModPhys.81.109}}

@article{Grigorenko2012NatPhot,
  author={A. N. Grigorenko and M. Polini and K. S. Novoselov}, title={Graphene plasmonics},
  journal={Nat. Photonics}, volume={6}, pages={749--758}, year={2012},
  doi={10.1038/nphoton.2012.262}}

@article{Crassee2012NanoLett,
  author={I. Crassee and M. Orlita and M. Potemski and A. L. Walter and M. Ostler and T. Seyller and I. Gaponenko and J. Chen and A. B. Kuzmenko},
  title={Intrinsic Terahertz Plasmons and Magnetoplasmons in Large Scale Monolayer Graphene},
  journal={Nano Lett.}, volume={12}, pages={2470--2474}, year={2012},
  doi={10.1021/nl300572y}}

@article{Yan2012NanoLett,
  author={H. Yan and Z. Li and X. Li and W. Zhu and P. Avouris and F. Xia},
  title={Infrared Spectroscopy of Tunable {Dirac} Terahertz Magneto-Plasmons in Graphene},
  journal={Nano Lett.}, volume={12}, pages={3766--3771}, year={2012},
  doi={10.1021/nl3016335}}

@article{Bandurin2022NatPhys,
  author={D. A. Bandurin and E. M{\"o}nch and K. Kapralov and I. Y. Phinney and K. Lindner and S. Liu and J. H. Edgar and I. A. Dmitriev and P. Jarillo-Herrero and D. Svintsov and S. D. Ganichev},
  title={Cyclotron resonance overtones and near-field magnetoabsorption via terahertz {Bernstein} modes in graphene},
  journal={Nature Physics}, volume={18}, pages={462--467}, year={2022},
  doi={10.1038/s41567-021-01494-8}}

@article{Monch2020NanoLett,
  author={E. M{\"o}nch and D. A. Bandurin and I. A. Dmitriev and I. Y. Phinney and I. Yahniuk and T. Taniguchi and K. Watanabe and P. Jarillo-Herrero and S. D. Ganichev},
  title={Observation of Terahertz-Induced Magnetooscillations in Graphene},
  journal={Nano Lett.}, volume={20}, pages={5943--5950}, year={2020},
  doi={10.1021/acs.nanolett.0c01918}}

@article{Moiseenko2025,
  author={I. Moiseenko and E. M{\"o}nch and K. Kapralov and D. A. Bandurin and S. D. Ganichev and D. Svintsov},
  title={Testing the tomographic {Fermi} liquid hypothesis with high-order cyclotron resonance},
  journal={Phys. Rev. Lett.}, volume={134}, pages={226902}, year={2025},
  doi={10.1103/zq1g-8d4s}}

@article{Yahniuk2025,
  author={I. Yahniuk and I. A. Dmitriev and A. L. Shilov and E. M{\"o}nch and M. Marocko and J. Eroms and D. Weiss and P. Sadovyi and B. Sadovyi and I. Grzegory and W. Knap and J. Gumenjuk-Sichevska and J. Wunderlich and D. A. Bandurin and S. D. Ganichev},
  title={Strongly nonlinear {Bernstein} modes in graphene reveal plasmon-enhanced near-field magnetoabsorption},
  journal={Phys. Rev. B}, volume={113}, pages={125418}, year={2026},
  doi={10.1103/32bf-v28g}}

@misc{DLMF,
  author={F. W. J. Olver and A. B. {Olde Daalhuis} and D. W. Lozier and B. I. Schneider and R. F. Boisvert and C. W. Clark and B. R. Miller and B. V. Saunders and H. S. Cohl and M. A. McClain},
  title={{NIST} Digital Library of Mathematical Functions, Release 1.2.6 of 2026-03-15},
  note={\url{https://dlmf.nist.gov/}}, year={2026}}

@article{Dmitriev2012RMP,
  author={I. A. Dmitriev and A. D. Mirlin and D. G. Polyakov and M. A. Zudov},
  title={Nonequilibrium phenomena in high {Landau} levels},
  journal={Rev. Mod. Phys.}, volume={84}, pages={1709--1763}, year={2012},
  doi={10.1103/RevModPhys.84.1709}}

@book{Watson1944,
  author={G. N. Watson}, title={A Treatise on the Theory of {Bessel} Functions},
  edition={2nd}, publisher={Cambridge University Press}, year={1944}}

@article{Hwang2007PRB,
  author={E. H. Hwang and S. {Das Sarma}}, title={Dielectric function, screening, and plasmons in two-dimensional graphene},
  journal={Phys. Rev. B}, volume={75}, pages={205418}, year={2007},
  doi={10.1103/PhysRevB.75.205418}}

@article{Wunsch2006NJP,
  author={B. Wunsch and T. Stauber and F. Sols and F. Guinea}, title={Dynamical polarization of graphene at finite doping},
  journal={New J. Phys.}, volume={8}, pages={318}, year={2006},
  doi={10.1088/1367-2630/8/12/318}}

@article{Monch2022PRB,
  author={E. M{\"o}nch and P. Euringer and G.-M. H{\"u}ttner and I. A. Dmitriev and D. Schuh and M. Marocko and J. Eroms and D. Bougeard and D. Weiss and S. D. Ganichev},
  title={Circular polarization immunity of the cyclotron resonance photoconductivity in two-dimensional electron systems},
  journal={Phys. Rev. B}, volume={106}, pages={L161409}, year={2022},
  doi={10.1103/PhysRevB.106.L161409}}

@article{BistrMac2009,
  author={R. Bistritzer and A. H. MacDonald}, title={Electronic Cooling in Graphene},
  journal={Phys. Rev. Lett.}, volume={102}, pages={206410}, year={2009},
  doi={10.1103/PhysRevLett.102.206410}}

@article{SongReizer2012,
  author={J. C. W. Song and M. Y. Reizer and L. S. Levitov}, title={Disorder-Assisted Electron-Phonon Scattering and Cooling Pathways in Graphene},
  journal={Phys. Rev. Lett.}, volume={109}, pages={106602}, year={2012},
  doi={10.1103/PhysRevLett.109.106602}}

@article{Monch2023PRB,
  author={E. M{\"o}nch and S. O. Potashin and K. Lindner and I. Yahniuk and L. E. Golub and V. Y. Kachorovskii and V. V. Bel'kov and R. Huber and K. Watanabe and T. Taniguchi and J. Eroms and D. Weiss and S. D. Ganichev},
  title={Cyclotron and magnetoplasmon resonances in bilayer graphene ratchets}, journal={Phys. Rev. B}, volume={107}, pages={115408}, year={2023},
  doi={10.1103/PhysRevB.107.115408}}

@article{Afanasiev2023PRB,
  author={A. N. Afanasiev and P. S. Alekseev and A. A. Greshnov and M. A. Semina}, title={Shear {Bernstein} modes in a two-dimensional electron liquid},
  journal={Phys. Rev. B}, volume={108}, pages={235124}, year={2023},
  doi={10.1103/PhysRevB.108.235124}}

@article{Newberger1982,
  author={B. S. Newberger}, title={New sum rule for products of {Bessel} functions with application to plasma physics},
  journal={J. Math. Phys.}, volume={23}, pages={1278--1281}, year={1982},
  doi={10.1063/1.525510}}

@misc{Tierz2025,
  author={M. Tierz},
  title={Amplitudes of {Hall} field-induced resistance oscillations with a two-harmonic density of states},
  eprint={2604.15700},
  archivePrefix={arXiv},
  primaryClass={cond-mat.mes-hall},
  year={2026}}

@article{Goerbig2011RMP,
  author={M. O. Goerbig}, title={Electronic properties of graphene in a strong magnetic field},
  journal={Rev. Mod. Phys.}, volume={83}, pages={1193--1243}, year={2011},
  doi={10.1103/RevModPhys.83.1193}}

@article{Fei2012Nature,
  author={Z. Fei and A. S. Rodin and G. O. Andreev and W. Bao and A. S. McLeod and M. Wagner and L. M. Zhang and Z. Zhao and M. Thiemens and G. Dominguez and M. M. Fogler and A. H. {Castro Neto} and C. N. Lau and F. Keilmann and D. N. Basov},
  title={Gate-tuning of graphene plasmons revealed by infrared nano-imaging},
  journal={Nature}, volume={487}, pages={82--85}, year={2012},
  doi={10.1038/nature11253}}

@article{Chen2012Nature,
  author={J. Chen and M. Badioli and P. Alonso-Gonz{\'a}lez and S. Thongrattanasiri and F. Huth and J. Osmond and M. Spasenovi{\'c} and A. Centeno and A. Pesquera and P. Godignon and A. Zurutuza Elorza and N. Camara and F. J. Garc{\'i}a de Abajo and R. Hillenbrand and F. H. L. Koppens},
  title={Optical nano-imaging of gate-tunable graphene plasmons},
  journal={Nature}, volume={487}, pages={77--81}, year={2012},
  doi={10.1038/nature11254}}

@article{HwangHuDasSarma2007,
  author={E. H. Hwang and B. Y.-K. Hu and S. {Das Sarma}},
  title={Inelastic carrier lifetime in graphene},
  journal={Phys. Rev. B}, volume={76}, pages={115434}, year={2007},
  doi={10.1103/PhysRevB.76.115434}}

@article{JiangShi2016NatMat,
  author={L. Jiang and Z. Shi and B. Zeng and S. Wang and J.-H. Kang and T. Joshi and C. Jin and L. Ju and J. Kim and T. Lyu and Y.-R. Shen and M. Crommie and H.-J. Gao and F. Wang},
  title={Soliton-dependent plasmon reflection at bilayer graphene domain walls},
  journal={Nat. Mater.}, volume={15}, pages={840--844}, year={2016},
  doi={10.1038/nmat4653}}

@article{ZhangFan2022PRL,
  author={H. Zhang and X. Fan and D. Wang and D. Zhang and X. Li and C. Zeng},
  title={Electric Field-Controlled Damping Switches of Coupled {Dirac} Plasmons},
  journal={Phys. Rev. Lett.}, volume={129}, pages={237402}, year={2022},
  doi={10.1103/PhysRevLett.129.237402}}

@article{SitenkoStepanov1957,
  author={A. G. Sitenko and K. N. Stepanov},
  title={On the Oscillations of an Electron Plasma in a Magnetic Field},
  journal={Sov. Phys. JETP}, volume={4}, pages={512--520}, year={1957}}

@article{ChaplikHeitmann1985,
  author={A. V. Chaplik and D. Heitmann},
  title={Geometric Resonances of Two-Dimensional Magnetoplasmons},
  journal={J. Phys. C: Solid State Phys.}, volume={18}, pages={3357--3363}, year={1985},
  doi={10.1088/0022-3719/18/17/012}}

@article{BatkeHeitmannTu1986,
  author={E. Batke and D. Heitmann and C. W. Tu},
  title={Plasmon and magnetoplasmon excitation in two-dimensional electron space-charge layers on {GaAs}},
  journal={Phys. Rev. B}, volume={34}, pages={6951--6960}, year={1986},
  doi={10.1103/PhysRevB.34.6951}}

@article{Bangert1996,
  author={D. E. Bangert and R. J. Stuart and H. P. Hughes and D. A. Ritchie and J. E. F. Frost},
  title={{Bernstein} modes in grating-coupled {2DEGs}},
  journal={Semicond. Sci. Technol.}, volume={11}, pages={352--359}, year={1996},
  doi={10.1088/0268-1242/11/3/013}}

@article{Bandurin2018NatComm,
  author={D. A. Bandurin and D. Svintsov and I. Gayduchenko and S. G. Xu and A. Principi and M. Moskotin and I. Tretyakov and D. Yagodkin and S. Zhukov and T. Taniguchi and K. Watanabe and I. V. Grigorieva and M. Polini and G. N. Goltsman and A. K. Geim and G. Fedorov},
  title={Resonant terahertz detection using graphene plasmons},
  journal={Nat. Commun.}, volume={9}, pages={5392}, year={2018},
  doi={10.1038/s41467-018-07848-w}}

@article{Dorozhkin2021JETPL,
  author={S. I. Dorozhkin and A. A. Kapustin and V. Umansky and J. H. Smet},
  title={Absorption of Microwave Radiation by Two-Dimensional Electron Systems Associated with the Excitation of Dimensional {Bernstein} Mode Resonances},
  journal={JETP Lett.}, volume={113}, pages={670--675}, year={2021},
  doi={10.1134/S0021364021100064}}

@article{Yavorskiy2025PRB,
  author={D. Yavorskiy and F. Le Mardel{\'e} and I. Mohelsky and M. Orlita and Z. Adamus and T. Wojtowicz and J. Wr{\'o}bel and K. Karpierz and J. {\L}usakowski},
  title={Low-energy excitations in multiple modulation-doped {CdTe}/({CdMg}){Te} quantum wells},
  journal={Phys. Rev. B}, volume={111}, pages={125414}, year={2025},
  doi={10.1103/PhysRevB.111.125414}}

\appendix

\section{Longitudinal dielectric function and factorized absorption}\label{app:bessel}

\paragraph*{A.1.\ Longitudinal conductivity from the kinetic equation.}
The quasiclassical nonlocal conductivity of a 2DEG in a perpendicular magnetic field is obtained from the linearized Boltzmann equation~\cite{Kapralov2022PRB,Dmitriev2012RMP}.
In the ballistic regime ($\omega\tau_p\gg 1$), the longitudinal conductivity for $\mathbf{q}\|\hat{x}$ is proportional to $Y_{00}^{(2)}$~\cite{Kapralov2022PRB}:
\begin{equation}
\sigma_{xx}(q,\omega) \propto \frac{-i}{\omega}\;Y_{00}^{(2)}(qR_c,\,\omega/\omega_c),
\label{eq:sigma-xx}
\end{equation}
where $R_c=v_F/\omega_c$ is the cyclotron radius and
\begin{equation}
Y_{00}^{(k)}(\zeta,\mu)
\equiv \sum_{s=-\infty}^{\infty}
\biggl(\frac{s}{\zeta}\biggr)^{\!k}
\frac{J_s^2(\zeta)}{1 - s/\mu + i0^+},
\label{eq:Ydef}
\end{equation}
with $\zeta\equiv qR_c$ and $\mu\equiv\omega/\omega_c$. The superscript $(k)$ in $Y_{00}^{(k)}$ denotes the power of $(s/\zeta)$ in the numerator. The object entering the longitudinal conductivity is $Y_{00}^{(2)}$, \emph{not} $Y_{00}^{(0)}$; the extra factor $(s/\zeta)^2$ arises from the kinematic relation between current and density through the continuity equation. (The exact overall coefficient of Eq.~\eqref{eq:sigma-xx} depends on the normalization conventions of Ref.~\cite{Kapralov2022PRB}; we fix it below via the long-wavelength magnetoplasmon limit, which is convention-independent.)

\paragraph*{A.2.\ Dielectric function.}
With gate screening at distance $d$ and the Coulomb kernel~\cite{Hwang2007PRB,Wunsch2006NJP}
\begin{equation}
V(q) = \frac{2\pi e^2}{\kappa q}\big(1-e^{-2qd}\big),
\label{eq:Vq}
\end{equation}
the longitudinal dielectric function is
\begin{equation}
\varepsilon_L(q,\omega) = 1 - \frac{2\,\omega_p^2(q)}{\omega^2}\;Y_{00}^{(2)}(\zeta,\mu),
\label{eq:epsL-Y}
\end{equation}
where
\begin{equation}
\omega_p^2(q) \equiv \frac{n_0\,q^2\,V(q)}{m_c},
\label{eq:omegap}
\end{equation}
is the standard 2D plasma frequency~\cite{Hwang2007PRB,Wunsch2006NJP}. The factor of $2$ and the sign in Eq.~\eqref{eq:epsL-Y} are fixed uniquely by the requirement that, as $qR_c\to 0$,
\[
Y_{00}^{(2)}\to\frac{\omega^2}{2(\omega^2-\omega_c^2)},
\]
so that $\varepsilon_L\to 1-\omega_p^2/(\omega^2-\omega_c^2)$, recovering the standard long-wavelength gated magnetoplasmon dispersion $\omega^2=\omega_c^2+\omega_p^2(q)$~\cite{Kapralov2022PRB}. An independent check is provided by the random-phase approximation (RPA) dielectric function of Volkov and Zabolotnykh~\cite{Volkov2014PRB} [their Eq.~(3)]: expanding their Bessel sum near the $n$th harmonic pole reproduces the same pole residue as Eq.~\eqref{eq:epsL-Y} with $m_c$ replacing the parabolic mass $m$.

\paragraph*{A.3.\ Relationship to the Newberger identity.}
The standard Bessel-product sums
\begin{equation}
S_{pq}(\mu,\zeta) \equiv \sum_{s=-\infty}^{\infty}
\frac{J_{s+p}(\zeta)\,J_{s+q}(\zeta)}{s-\mu},
\label{eq:Spqdef}
\end{equation}
close via contour integrals~\cite{Newberger1982,Tierz2025,Watson1944} into
\begin{equation}
S_{00}(\mu,\zeta) = -\frac{\pi}{\sin(\pi\mu)}\,J_{-\mu}(\zeta)\,J_{\mu}(\zeta).
\label{eq:S00}
\end{equation}
$Y_{00}^{(k)}$ is related to $S_{00}$ by exact algebraic identities. In particular, using $\sum_s J_s^2=1$ and $\sum_s s\,J_s^2=0$~\cite{Watson1944,DLMF}:
\begin{align}
Y_{00}^{(0)} &= -\mu\,S_{00},
\label{eq:Y0-S}\\
Y_{00}^{(2)} &= -\frac{\mu^2}{\zeta^2}\bigl(1 + \mu\,S_{00}\bigr).
\label{eq:Y2-S}
\end{align}
Near $\mu=n$, $S_{00}$ has a pole with residue $-J_n^2(\zeta)$. Thus
\begin{equation}
Y_{00}^{(2)} \;\xrightarrow{\;\mu\to n\;}\;
\frac{n^3}{\zeta^2}\,\frac{J_n^2(\zeta)}{\mu - n} + [\text{smooth}].
\label{eq:Y2-pole}
\end{equation}

\paragraph*{A.4.\ Resonant dielectric function and the splitting $\Omega_n$.}
Substituting Eq.~\eqref{eq:Y2-pole} into Eq.~\eqref{eq:epsL-Y} and converting $\mu-n=\Delta_\omega/\omega_c$, where $\Delta_\omega\equiv\omega-n\omega_c$:
\begin{equation}
\varepsilon_L(q,\omega) \simeq 1 - \frac{\Omega_n(q)}{\Delta_\omega + i\Gamma_n},
\label{eq:epsL-res}
\end{equation}
with the splitting
\begin{equation}
\Omega_n(q) = \frac{2\,\omega_p^2(q)\,n\omega_c}{\omega^2}\;h_n(\zeta),
\qquad
h_n(\zeta) \equiv \Bigl(\frac{n}{\zeta}\Bigr)^{\!2} J_n^2(\zeta).
\label{eq:Omega-def}
\end{equation}
At fixed $\omega$ and to leading order in $\Omega_n/\omega$: $n\omega_c\simeq\omega$, $q_n^{(0)}\simeq\omega/v_F$, $R_c\simeq nv_F/\omega$, and $h_n(n)=J_n^2(n)$. The entire prefactor of $h_n$ is then $n$-independent.

\paragraph*{A.5.\ BM dispersion and curvature.}
Setting $\varepsilon_L=0$ in the weak-coupling limit gives $\omega_{\rm BM}(q) = n\omega_c + \Omega_n(q)$~\cite{Roldan2011PRB,Dmitriev2012RMP}. Writing $\Omega_n(q)=A(q)\,h_n(qR_c)$ with $A(q)\equiv 2\omega_p^2(q)n\omega_c/\omega^2$, the full curvature at the physical turning point is $\mathcal{K}_n^* = |A''h_n + 2A'R_c h_n' + A R_c^2 h_n''|_{q_n^*}$. The \emph{baseline} curvature, evaluated at the Bessel peak $\zeta_n^{(h)}$ where $h_n'=0$, is
\begin{equation}
\mathcal{K}_n^{(0)}
\equiv A(k_\omega)\,R_c^2\;\big|h_n''(\zeta_n^{(h)})\big|,
\qquad A(k_\omega)=\frac{2\,\omega_p^2(k_\omega)\,n\omega_c}{\omega^2},
\label{eq:Kn-full}
\end{equation}
where $A(k_\omega)$ is common to all harmonics at fixed $\omega$ (since $n\omega_c=\omega$) and therefore cancels from $\mathcal B^{(0)}_{nm}$. At fixed $\omega$, $R_c^2=n^2v_F^2/\omega^2\propto n^2$ while $|h_n''|\propto n^{-4/3}$, giving $\mathcal{K}_n\propto n^{2/3}$.

\paragraph*{A.6.\ Turning-point projection.}
Near the turning point, expand $\Omega_n(q)\simeq\Omega_n(q_n^*)-\tfrac12\mathcal{K}_n(q-q_n^*)^2$ and evaluate the $q$-integral in the narrow-window approximation~\cite{Bandurin2022NatPhys}. The peak absorption takes the factorized form~\eqref{eq:Lambda}, with the $n$-scaling at fixed $\omega$:

\vspace{4pt}

\begin{center}
\begin{tabular}{lll}
\hline
Factor & & $n$-scaling \\
\hline
$|D(q_n^*)|^2$ & launcher & $n^0$ \\
$\Omega_n(q_n^*)$ & splitting at turning point & $n^{-2/3}$ \\
$\mathcal{K}_n^{-1/2}$ & turning-point enhancement & $n^{-1/3}$ \\
$\Gamma_n^{-1/2}$ & linewidth & $n^0$ \\
\hline
$\Lambda_n$ & net & $n^{-1}$ \\
\hline
\end{tabular}
\end{center}
The inter-harmonic baseline is $\mathcal{B}_{nm}^{(0)}\simeq m/n$.

\paragraph*{A.7.\ Turning-point asymptotics.}
The Bessel function $J_n(\zeta)$ oscillates for $\zeta > n$ and decays for $\zeta < n$. In the regime $\zeta = n + s\,n^{1/3}$ with $|s| = \mathcal{O}(1)$, the uniform Airy expansion given in the NIST Digital Library of Mathematical Functions (DLMF) \S10.20~\cite{DLMF} gives
\begin{equation}
J_n(\zeta)\sim \frac{2^{1/3}}{n^{1/3}}\,\operatorname{Ai}\big(-2^{1/3}s\big).
\label{eq:airyJ}
\end{equation}
At $\zeta = n$: $J_n^2(n)\sim 2^{2/3}\operatorname{Ai}^2(0)\,n^{-2/3}$, yielding the result quoted in Eq.~\eqref{eq:Jn2}.

\paragraph*{A.8.\ Launcher model.}
For the synthetic tests we use the minimal smooth launcher proxy
\begin{equation}
|D(q)|^2 \propto \frac{(1-e^{-2qd})}{[1+(q\ell)^2]^2},
\label{eq:Dq}
\end{equation}
which is not intended to reproduce the full contact-diffraction calculation of Ref.~\cite{Bandurin2022NatPhys}, where finite contact geometry, substrate response, and angular structure of the near field enter explicitly. The proxy retains only the features needed for the ratio test: suppression at very small $q$ (through the same gate-distance $d$ that enters the Coulomb kernel~\eqref{eq:Vq}), a broad maximum near $q\sim 1/\ell$, and a smooth high-$q$ roll-off. A device-specific analysis should replace $D(q)$ by the appropriate angle-averaged longitudinal near-field spectrum. The phrase ``unscreened Coulomb'' used below refers to the in-plane Coulomb interaction $V(q)=2\pi e^2/(\kappa q)$ between electrons in the 2DEG, and is a separate choice from the launcher's near-field shape: when we quote a launcher $d/\ell$ in the ``unscreened'' case, it is the launcher's gate-distance parameter that sets the small-$q$ cutoff of $|D|^2$, while $V(q)$ itself is taken unscreened in the dielectric function.

\paragraph*{A.9.\ Assumptions and their corrections.}
For reference, Table~\ref{tab:validity} lists the assumptions entering the peak formula~\eqref{eq:Lambda} and the leading correction when each is relaxed.

\begin{table}[!htbp]
\caption{Assumptions entering the peak formula and the leading correction when each is relaxed.}
\label{tab:validity}
\begin{ruledtabular}
\begin{tabular}{p{0.30\columnwidth}p{0.62\columnwidth}}
Assumption & If relaxed \\
\hline
Quasiclassical ($E_F\!\gg\!\hbar\omega_c$, high Landau-level filling) &
  Landau-quantization corrections; BM branch structure modified~\cite{Roldan2011PRB,Goerbig2011RMP} \\[2pt]
Ballistic ($\omega\tau_{ee}\!\gg\!1$) &
  Hydrodynamic damping replaces Bessel structure~\cite{Kapralov2022PRB} \\[2pt]
Broadband launcher &
  Factorization receives derivative corrections; angular patterns become $n$-dependent \\[2pt]
Narrow turning window &
  Finite-window integration corrections to $\mathcal{K}_n^{-1/2}$ lineshape \\[2pt]
Common $\Gamma_n$ &
  Explicit factor $\sqrt{\Gamma_{m}/\Gamma_{n}}$ in ratios~\cite{Moiseenko2025} \\[2pt]
Common $\beta_n$ &
  $\widetilde{\mathcal R}^{\rm eff}_{nm}$ picks up known factor $\beta_n/\beta_m$ \\[2pt]
Shared cooling geometry &
  Onset relation acquires $\mathcal W_n/\mathcal W_m\neq 1$ factor \\[2pt]
Fixed-$\omega$ comparison &
  If $\omega$ varies between harmonics, $q_n\neq q_m$ at leading order; full $q$-dependence must be retained \\
\end{tabular}
\end{ruledtabular}
\end{table}
\FloatBarrier

\section{Uncertainty propagation}\label{app:uncertainty}

For the reduced ratio inferred from low-power amplitudes using the common-$\Gamma_n$ baseline,
\begin{equation}
\mathcal R^{\rm res}_{nm}
=
\frac{\widehat A_n/\widehat A_m}{\mathcal B^{(0)}_{nm}},
\label{eq:Rtilde-def}
\end{equation}
we have $\ln \mathcal R^{\rm res}_{nm} = \ln \widehat A_n - \ln \widehat A_m - \ln \mathcal B^{(0)}_{nm}$. To linear order,
\begin{equation}
\sigma^2_{\ln \mathcal R^{\rm res}_{nm}}
=
\biggl(\frac{\sigma_{A_n}}{A_n}\biggr)^{\!2}+
\biggl(\frac{\sigma_{A_m}}{A_m}\biggr)^{\!2}+
\sigma^2_{\ln \mathcal B^{(0)}_{nm}},
\label{eq:logerr}
\end{equation}
and the multiplicative error bar is $\sigma_{\mathcal R^{\rm res}_{nm}} = \mathcal R^{\rm res}_{nm}\,\sigma_{\ln \mathcal R^{\rm res}_{nm}}$. If the full linewidth-curvature form~\eqref{eq:ratio-general} is used instead of the common-$\Gamma_n$ baseline, the corresponding $\mathcal K_n$ and $\Gamma_n$ terms enter in quadrature with prefactor $1/4$. For ratios sharing the same fitted amplitudes, the covariance matrix $C_{ij}=\mathrm{Cov}(\ln A_i,\ln A_j)$ should be retained:
\begin{equation}
\mathrm{Var}\!\bigl[\ln \mathcal R^{\rm res}_{nm}\bigr]
=
C_{nn}+C_{mm}-2C_{nm}+\sigma^2_{\ln \mathcal B^{(0)}_{nm}}.
\label{eq:cov}
\end{equation}

\section{Published-data comparison}\label{app:published-data}

As an illustrative comparison with published data, not a reanalysis of raw data, the published power-sweep data of Yahniuk \emph{et al.}~\cite{Yahniuk2025} (their Appendix Fig.~A.3, $f=2.54$~THz, $T=1.8$~K, $n_e=2.83\times 10^{12}\,\mathrm{cm}^{-2}$) yield $A_2\approx 0.44\pm 0.05\,\Omega\,\mathrm{cm}^2/\mathrm{W}$ from the plotted points with $I_{s,2}\approx 0.65\,\mathrm{W/cm}^2$. For the $n=3$ slope, two estimates disagree: direct figure digitization gives $A_3^{\rm(dig)}\approx 0.15\pm 0.02$, whereas the empirical relation $A_n\cdot I_{s,n}\approx\text{const}$ (Sec.~\ref{sec:nonlinear}), applied to the same data set with $I_{s,3}\approx 1.15\,\mathrm{W/cm}^2$, gives $A_3^{\rm(prod)}=A_2\,I_{s,2}/I_{s,3}\approx 0.25$, a factor of $1.7$ larger. Using $\mathcal B^{(0)}_{23}=1.54$, the effective reduced ratio is $\widetilde{\mathcal R}^{\rm res}_{23}\approx 1.9\pm 0.3$ from the digitized $A_3$ and $\approx 1.15\pm 0.2$ from the $A\cdot I_s$ construction. The latter, however, is \emph{not} an independent check: because $A_3^{\rm(prod)}=A_2 I_{s,2}/I_{s,3}$ by construction, $\widetilde{\mathcal R}^{\rm res}_{23}$ evaluated with that $A_3$ reduces to $(A_2/A_3^{\rm(prod)})/\mathcal B^{(0)}_{23} = (I_{s,3}/I_{s,2})/\mathcal B^{(0)}_{23} = \widetilde{\mathcal S}_{23}^{-1}$, where $\widetilde{\mathcal S}_{23}\equiv (I_{s,2}/I_{s,3})\,\mathcal B^{(0)}_{23}$. Its ``agreement'' with the nonlinear onset quantity $\widetilde{\mathcal S}_{23}=0.87$ is therefore a consistency construction, not a test; only the directly digitized estimate $\widetilde{\mathcal R}^{\rm res}_{23}\approx 1.9\pm 0.3$ carries independent information. Taken at face value, that value is above the $m/n$ baseline, consistent with a smooth $\mathcal G_n^{\rm int}$ that decreases with~$n$; with $\tau_2/\tau_3\sim 1.5$--$2$ from Ref.~\cite{Moiseenko2025} the implied linewidth factor $\sqrt{\Gamma_3/\Gamma_2}\approx 1.2$--$1.4$ further reduces $\widetilde{\mathcal R}^{\rm res}_{23}$ by $20$--$30\%$. A purpose-built multi-harmonic measurement at one frequency, as set up in the synthetic recovery test of Sec.~\ref{sec:ratios}, would convert this factor-of-$1.7$ ambiguity into an unambiguous residue extraction.

\FloatBarrier
\vspace*{0.8\baselineskip}
\section{Misspecification stress tests}\label{app:adversarial}

\begin{figure*}[!tbp]
\centering
\includegraphics[width=0.48\textwidth]{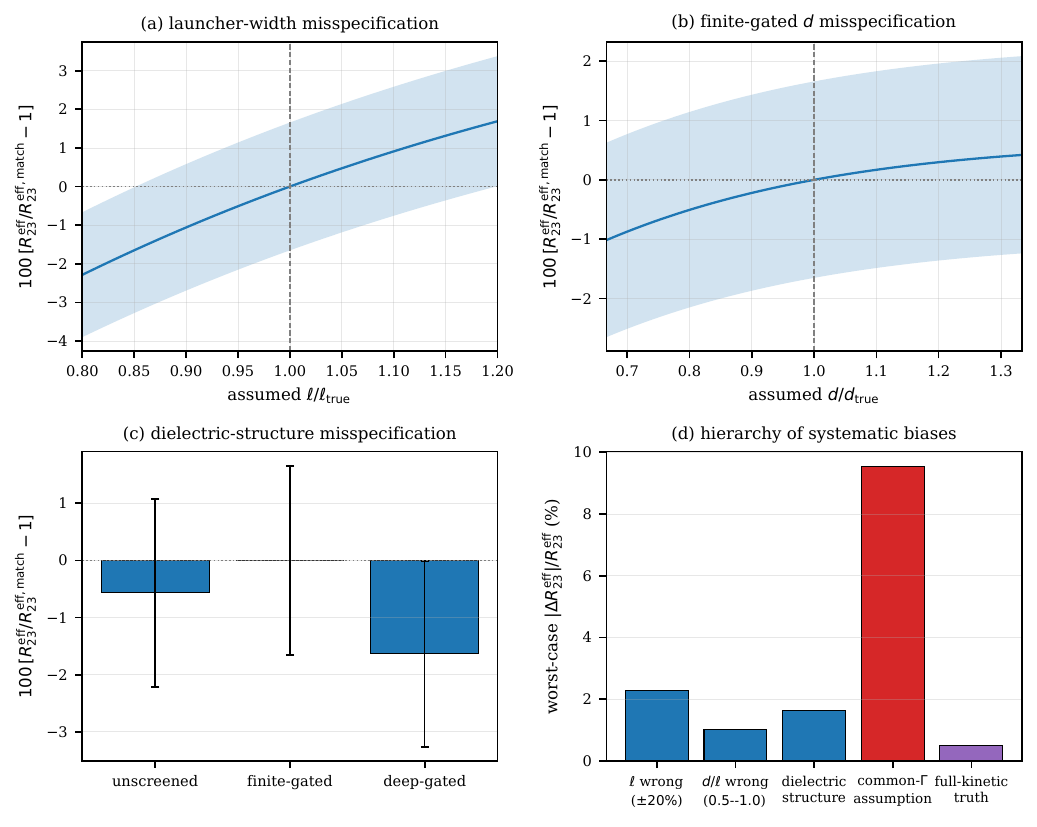}\hfill
\includegraphics[width=0.48\textwidth]{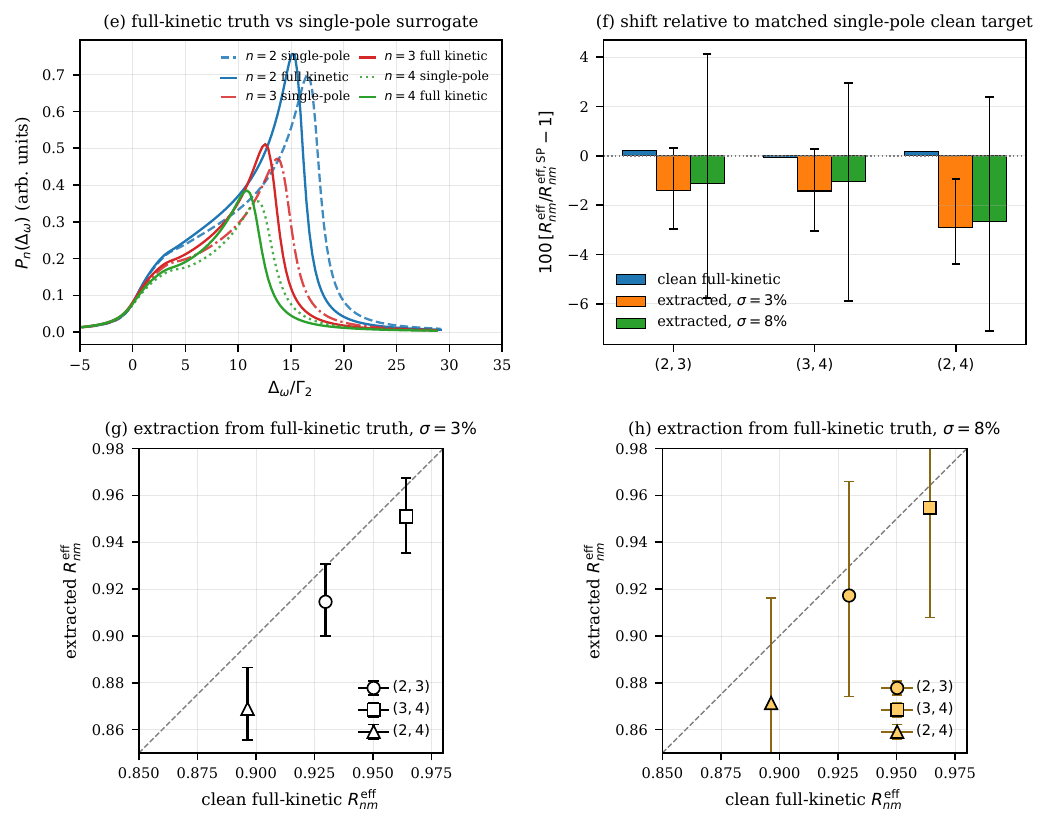}
\caption{\emph{Panels (a--d):} Misspecification stress tests for the $(n,m)=(2,3)$ pair. Data are generated from the full $q$-integral with the finite-gated generating model of Fig.~\ref{fig:recovery}; the extraction is repeated with incorrect nuisance assumptions. (a)~Bias when the launcher width $\ell$ differs from the generating value. (b)~Bias when $d$ is misspecified ($d_{\rm ass}/\ell=0.5$--$1.0$, i.e.\ $d_{\rm ass}/d_{\rm true}\approx 0.67$--$1.33$). (c)~Bias from the wrong dielectric structure. (d)~Worst-case biases, including the common-$\Gamma$ assumption and the full-kinetic data generator (panels e--h). Within this smooth benchmark family, geometric/dielectric misspecification induces $\sim 1$--$2\%$ shifts; the linewidth hierarchy dominates at $\sim 10\%$; the full-kinetic mismatch is sub-percent. --- \emph{Panels (e--h):} Residual generator$\,\neq\,$extractor test using the full Bessel-sum kinetic dielectric $\varepsilon_L=1-(2\omega_p^2/\omega^2)\,Y_{00}^{(2)}$ (Eq.~\eqref{eq:epsL-Y}, $|s|\le 60$) on the data-generating side, single-pole surrogate on the extraction side. (e)~Noiseless peaks: full-kinetic (solid) vs.\ single-pole (dashed/dot-dashed/dotted); the $\sim\!8$--$10\%$ amplitude uplift from non-resonant harmonic tails is visible. (f)~Noiseless corrected-ratio shifts: $+0.22\%$, $-0.06\%$, $+0.16\%$ for $(2,3)$, $(3,4)$, $(2,4)$. (g,\,h)~Extracted vs.\ noiseless full-kinetic $\widetilde{\mathcal R}^{\rm eff}_{nm}$ at $3\%$ and $8\%$ noise. The $1$--$3\%$ downward bias is a known lineshape-fit property. Broadening is phenomenological ($\mu\to\mu+i\Gamma_n/\omega_c$); the full collision-renormalized kernel is left for future work.}
\label{fig:adversarial}
\end{figure*}

As a complement to the matched-model recovery of Fig.~\ref{fig:recovery}, we repeat the $(n,m)=(2,3)$ extraction --- the most sensitive pair, because it has the largest pairwise turning-point mismatch --- with intentionally misspecified nuisance models. Data are generated from the full $q$-integral with the same nominal generating model as Fig.~\ref{fig:recovery}; the extraction is then repeated with wrong $\ell$, wrong $d$, a wrong dielectric form (unscreened, finite-gated, deep-gated), or a common-$\Gamma$ assumption. The resulting biases in the extracted $\widetilde{\mathcal R}^{\rm eff}_{23}$ are summarized in Fig.~\ref{fig:adversarial}. As a further test, we replace the generating-side dielectric with the full ballistic Bessel-sum kinetic function (Eq.~\eqref{eq:epsL-Y}) while keeping the extraction-side single-pole surrogate unchanged; the pairwise noiseless corrected-ratio shifts are sub-percent for all harmonic pairs [Fig.~\ref{fig:adversarial}(e--h)], confirming that the resonant-pole approximation is not the limiting systematic.

\end{document}